\documentclass[preprint,preprintnumbers, prd, floatfix, superscriptaddress,nofootinbib] {revtex4-1}
\usepackage{epsfig}
\usepackage{subfigure}
\usepackage{dcolumn}
\usepackage{bm}
\usepackage[usenames ,dvipsnames]{xcolor}
\usepackage{slashed}
\usepackage{graphicx,color}
\usepackage{lineno}

\begin{document}
\title{Topological $SU(3)_f$ approach for two-body $\Omega_c$ weak decays}

\author{Y.~L.~Wang}
\email{ylwang0726@163.com}
\affiliation{School of Physics and Information Engineering, Shanxi Normal University,
Taiyuan 030031, China}

\author{H.~J.~Zhao}
\email{hjzhao@163.com}
\affiliation{School of Physics and Information Engineering, Shanxi Normal University,
Taiyuan 030031, China}

\author{Y.~K.~Hsiao}
\email{yukuohsiao@gmail.com}
\affiliation{School of Physics and Information Engineering, Shanxi Normal University,
Taiyuan 030031, China}

\date{\today}

\begin{abstract}
We explore the two-body non-leptonic weak decays of $\Omega_c^0$ 
into final states ${\bf B}^{(*)}M$ and ${\bf B}^{(*)}V$, 
where ${\bf B}^{(*)}$ denotes an octet (a decuplet) baryon
and $M(V)$ represents a pseudoscalar (vector) meson. 
We employ the topological $SU(3)_f$ approach to depict 
and parameterize the $W$-emission and $W$-exchange processes. 
We find that the topological parameters can be associated and combined, 
making them extractable for calculation.
Consequently, we explain the partially measured branching fractions 
relative to ${\cal B}(\Omega_c^0\to \Omega^-\pi^+)$, 
recombined or kept as the following ratios: 
${\cal B}(\Omega_c^0\to\Xi^{*0}\bar K^{*0})/{\cal B}(\Omega_c^0\to\Omega^-\rho^+)=0.28\pm 0.11$, 
${\cal B}(\Omega_c^0\to\Xi^-\pi^+)/{\cal B}(\Omega_c^0\to\Xi^{0}\bar K^{0})=0.10\pm 0.02$, and 
${\cal B}(\Omega_c^0 \to \Omega^- K^+)/{\cal B}(\Omega_c^0 \to \Omega^- \pi^+)=0.06\pm 0.01$.
In particular, we present ${\cal B}(\Omega_c^0 \to \Xi^0 \pi^0)=(2.3\pm0.2)\times 10^{-4}$ 
as half the value of ${\cal B}(\Omega_c^0 \to \Xi^- \pi^+)$ in the approximate isospin relation, 
and highlight potential candidates for testing $SU(3)_f$ symmetry breaking.
\end{abstract}

\maketitle
\section{introduction}
\begin{figure}[t]
\includegraphics[width=1.9in]{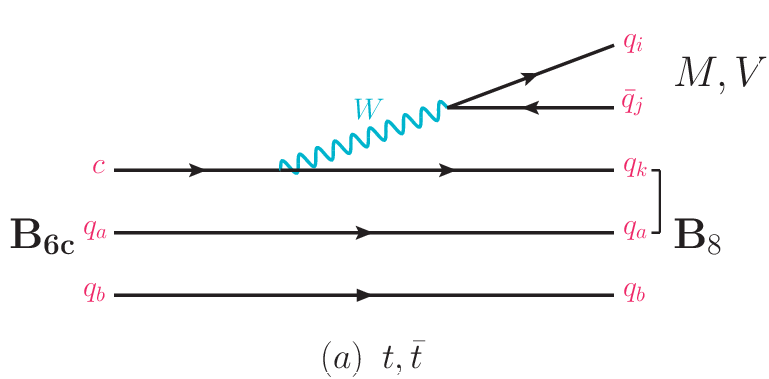}
\includegraphics[width=1.9in]{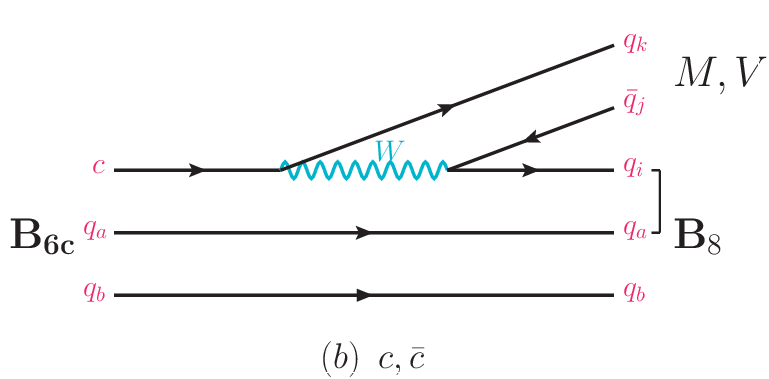}\\[2mm]
\includegraphics[width=1.9in]{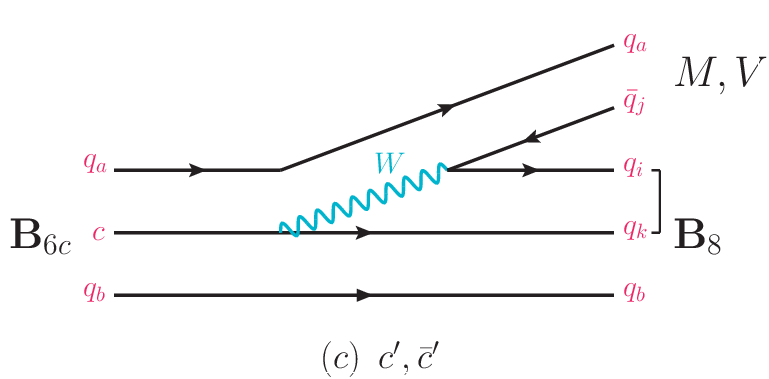}
\includegraphics[width=1.9in]{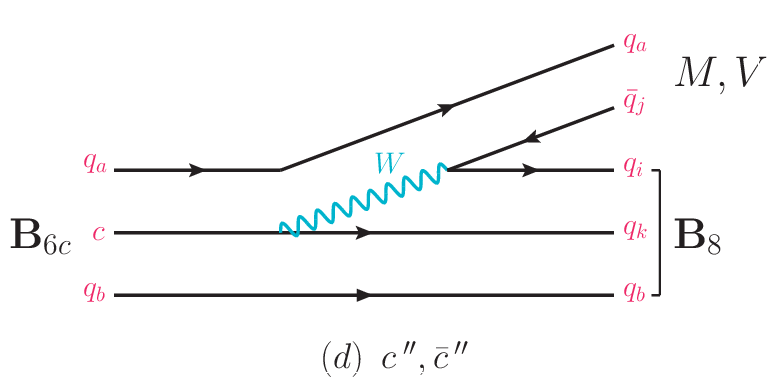}\\[2mm]
\includegraphics[width=1.9in]{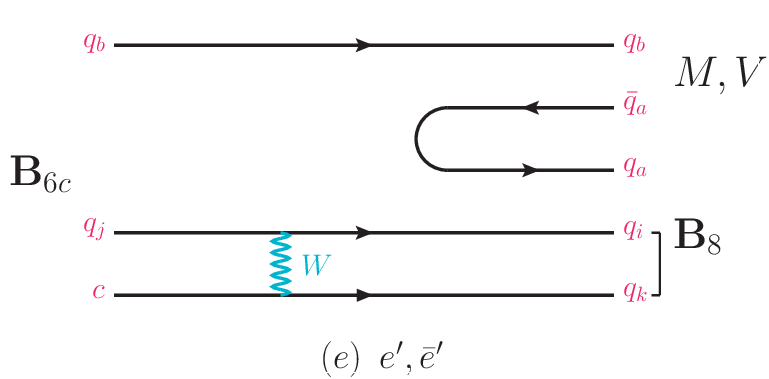}
\includegraphics[width=1.9in]{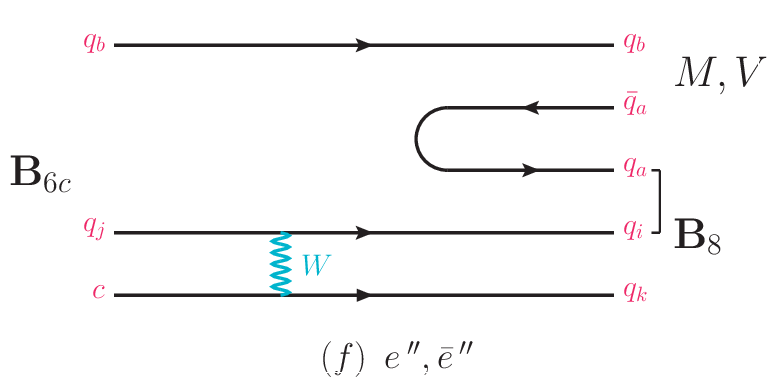}\\[2mm]
\includegraphics[width=1.9in]{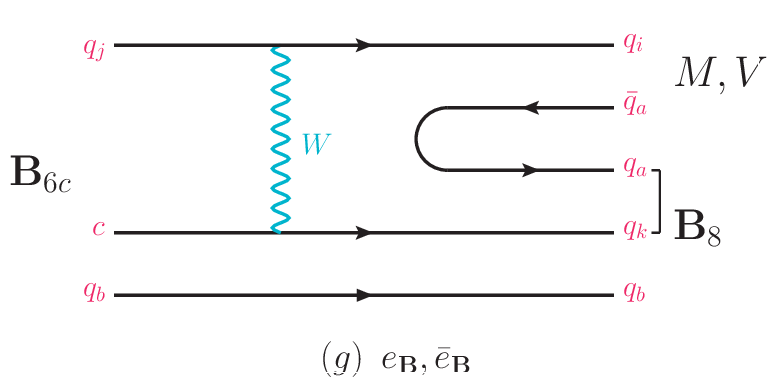}
\includegraphics[width=1.9in]{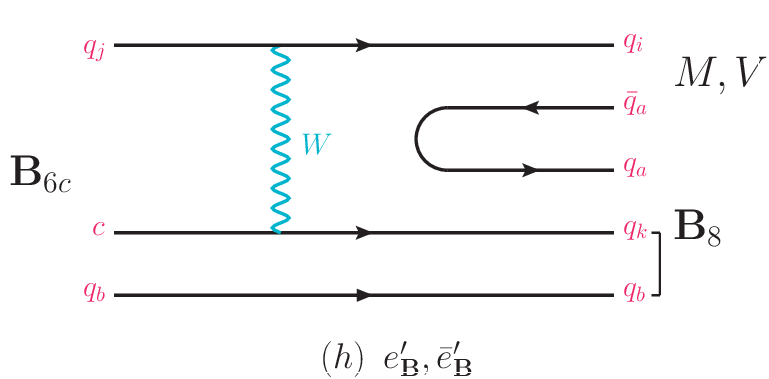}\\[2mm]
\includegraphics[width=1.9in]{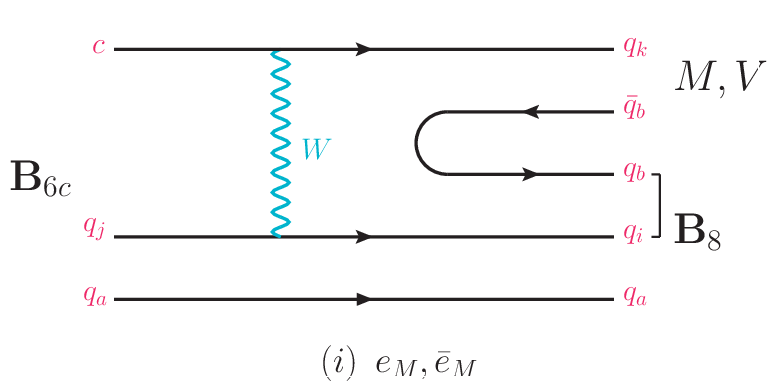}
\includegraphics[width=1.9in]{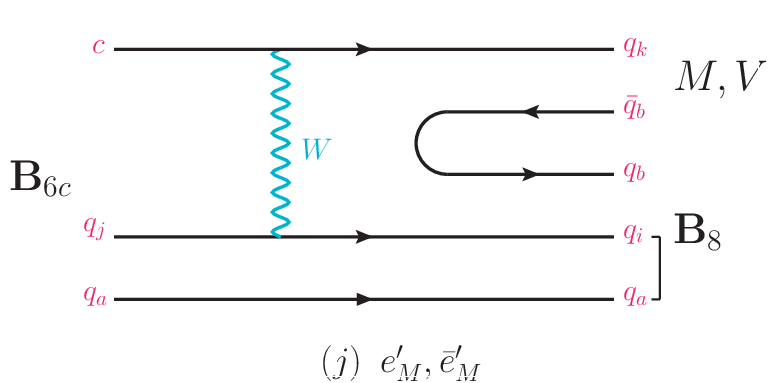}
\caption{Topological diagrams of ${\bf B}_{6c}\to {\bf B}_8 M(V)$ decays,
where the notations ``]'' denote the asymmetric quark orderings
of octet baryon states.}\label{fig1}
\end{figure}

The $\Omega_c^0(css)$ baryon, with $ss$ in a symmetric quark state, 
is unique as the only sextet charmed baryon (${\bf B}_{6c}$) that undergoes weak decays.
Similar to the anti-triplet charmed baryon counterparts, 
denoted as ${\bf B}_{3c}=(\Xi_c^0,\Xi_c^+,\Lambda_c^+)$,
the two-body non-leptonic weak decays of $\Omega_c^0$ 
occur through a variety of configurations, as represented 
by the $W$-boson emission and $W$-boson exchange topological diagrams 
in Figs.~\ref{fig1} and \ref{fig2}~\cite{Kohara:1991ug,Zhao:2018mov, 
Hsiao:2020iwc,Hsiao:2021nsc}. However, the contributions of 
these configurations to the total branching fractions 
are not well understood. In recent years, experimental collaborations 
such as ALICE~\cite{ALICE:2022cop, ALICE:2023sgl}, 
Belle~\cite{Yelton:2017uzv, Belle:2021dgc, Belle:2022yaq}, 
and LHCb~\cite{LHCb:2023fvd} have conducted reanalyses and measurements, 
which have opened a new window for extracting valuable information
regarding these configurations.

Due to the lower production rate of $\Omega_c^0$ and 
the unclear fragmentation fraction~\cite{ALICE:2023sgl}, 
absolute branching fractions of $\Omega_c^0$ decays have not been made available.
However, it is still possible to measure the rates of the branching fractions:
${\cal R}(\Omega_c^0\to{\bf B}^{(*)}X)\equiv 
{\cal B}(\Omega_c^0\to {\bf B}^{(*)}X)/{\cal B}(\Omega_c^0\to \Omega^-\pi^+)$.
Here, ${\bf B}^{(*)}$ denotes an octet (a decuplet) baryon, and 
$X$ represents a lepton pair, a pseudoscalar meson ($M$), a vector meson ($V$),
or a meson pair. These rates are reported as follows:
\begin{eqnarray}\label{data1}
&&
{\cal R}(\Omega_c^0\to\Omega^-\rho^+)
=1.80\pm 0.33~\text{(Belle, pdg)~\cite{Yelton:2017uzv,pdg}}\,,
\nonumber\\
&&
{\cal R}_e\equiv{\cal R}(\Omega_c^0\to\Omega^-e^+\nu_e)
=1.98\pm 0.15~\text{(Belle, pdg)~\cite{pdg,Belle:2021dgc}}\,,
\nonumber\\
&&
{\cal R}(\Omega_c^0\to\Xi^0 \bar K^0,\Xi^0 \bar K^{*0},\Xi^{*0} \bar K^{*0})
=(1.64\pm 0.29,1.02\pm 0.24,0.51\pm 0.17)~\text{(pdg)~\cite{pdg}}\,,
\nonumber\\
&&
{\cal R}_{\Xi\pi}\equiv{\cal R}(\Omega_c^0\to\Xi^- \pi^+)
=(16.1\pm 1.0)\times 10^{-2}~\text{(Belle, LHCb)~\cite{Belle:2022yaq,LHCb:2023fvd}}\,,\nonumber\\
&&
{\cal R}(\Omega_c^0\to\Omega^- K^+)
=(6.08\pm 0.51\pm 0.40)\times 10^{-2}~\text{(LHCb)~\cite{LHCb:2023fvd}}\,.
\end{eqnarray} 
In Eq.~(\ref{data1}), we use the resonant relations:
${\cal R}(\Omega_c^0\to\Omega^-\rho^+)=
{\cal R}(\Omega_c^0\to\Omega^-\pi^+\pi^0)/{\cal B}(\rho^+\to\pi^+\pi^0)$ and
${\cal R}(\Omega_c^0\to\Xi^{*0} \bar K^{*0})=
{\cal R}(\Omega_c^0\to\Xi^-\pi^+\bar K^{*0})/{\cal B}(\Xi^{*0}\to \Xi^-\pi^+)$. Additionally,
we calculate the weighted average of 
${\cal R}_{\Xi\pi}=(25.3\pm 5.3\pm 3.0)\times 10^{-2}$~\cite{Belle:2022yaq} and 
${\cal R}_{\Xi\pi}=(15.81\pm 0.87\pm 0.43\pm 0.16)\times 10^{-2}$~\cite{LHCb:2023fvd},
as measured by Belle and LHCb, respectively.

The decays of $\Omega_c^0$ have been extensively studied 
by various research groups~\cite{Perez-Marcial:1989sch,Xu:1992sw,
Cheng:1993gf,Cheng:1995fe,Pervin:2006ie,Dhir:2015tja,Gutsche:2018utw,Cheng:1996cs,
Zhao:2018zcb,Hsiao:2020gtc,Huang:2021ots,Aliev:2022gxi,Hu:2020nkg,Groote:2021pxt,Wang:2022zja}.
To calculate the branching fractions for $\Omega_c^0\to{\bf B}^{(*)}M$,
one can factorize the $W$-emission amplitudes,
as depicted in Fig.~\ref{fig1}$(a,b)$ and Fig.~\ref{fig2}$(a,b)$,
into two separate matrix elements~\cite{Hsiao:2020gtc,Hsiao:2021mlp},
${\cal M}\propto\langle M|(\bar q_1 q_2)|0\rangle\langle{\bf B}^{(*)}|(\bar q_3 c)|\Omega_c^0\rangle$,
with the latter being computed using quark models~\cite{Perez-Marcial:1989sch,Cheng:1993gf,
Cheng:1995fe,Pervin:2006ie,Zhao:2018zcb,Hsiao:2020gtc,Aliev:2022gxi}. 
In contrast, even an estimation of 
the non-factorizable amplitudes can be challenging, as depicted 
in Fig.~\ref{fig1}(c-j) and Fig.~\ref{fig2}(c-f).

Apart from the pole model~\cite{Cheng:1993gf,Hu:2020nkg},
which partially estimates the non-factorizable amplitudes,
it is worth noting that the $SU(3)$ flavor $[SU(3)_f]$ symmetry can 
address all contributions~\cite{Savage:1989qr,
Savage:1991wu,Sharma:1996sc,Lu:2016ogy,Geng:2017mxn, 
Geng:2017esc,Geng:2018plk,Geng:2018upx,Hsiao:2019yur,Huang:2021aqu,
Jia:2019zxi,Wang:2017gxe,Wang:2019dls,Wang:2022kwe,Geng:2018bow,
Kohara:1991ug,Chau:1995gk,Zhao:2018mov,Hsiao:2020iwc,Pan:2020qqo,
Chau:1995gk,Zeppenfeld:1980ex,He:2018php,He:2018joe,Wang:2020gmn,
Hsiao:2021nsc,Hsiao:2023mud,Xing:2023dni,Zhong:2022exp}. Meanwhile, 
it avoids the complexities of model calculations.
%
%
Thus, we propose the utilization of the $SU(3)_f$-based 
topological diagram approach (TDA)~\cite{Kohara:1991ug,Chau:1995gk,Zhao:2018mov,
Wang:2020gmn,Hsiao:2020iwc,Pan:2020qqo,Hsiao:2021nsc,Hsiao:2023mud}, 
which parameterizes topological diagrams and establish strict $SU(3)_f$ relations 
for $\Omega_c^0$ decays. We will also employ the irreducible $SU(3)_f$ approach 
(IRA)~\cite{Savage:1989qr,
Savage:1991wu,Sharma:1996sc,Lu:2016ogy,Geng:2017mxn,
Geng:2017esc,Geng:2018plk,Geng:2018upx,Hsiao:2019yur,Huang:2021aqu,Jia:2019zxi,
Zeppenfeld:1980ex,He:2018php,He:2018joe,Wang:2017gxe,Wang:2019dls,Wang:2022kwe,
Geng:2018bow,Huang:2021aqu,Xing:2023dni,Zhong:2022exp},
which aids in deriving approximate relations for the topological parameters.
This will enable us to explore sub-processes of Figs.~\ref{fig1}~and~\ref{fig2}
in the two-body non-leptonic $\Omega_c^0$ decays, specifically 
$\Omega_c^0\to{\bf B}^{(*)}M$ and $\Omega_c^0\to{\bf B}^{(*)}V$. Additionally, 
we will calculate branching fractions that can be experimentally measured 
at facilities such as Belle and LHCb.

\section{Formalism}
\begin{figure}[t]
\includegraphics[width=1.8in]{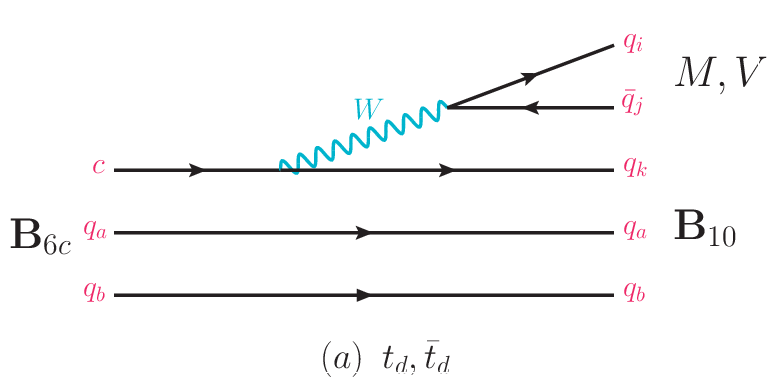}
\includegraphics[width=1.8in]{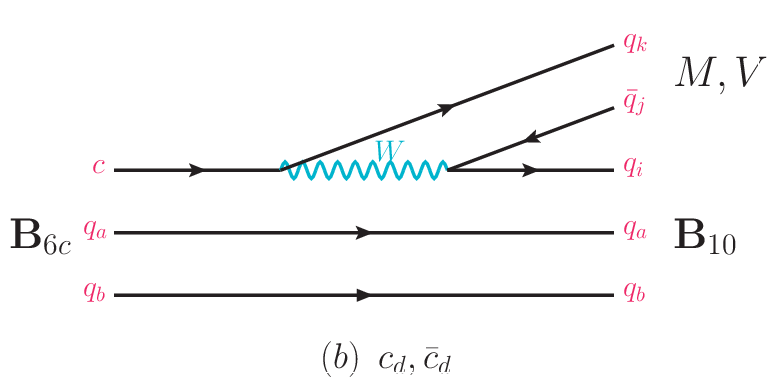}
\includegraphics[width=1.8in]{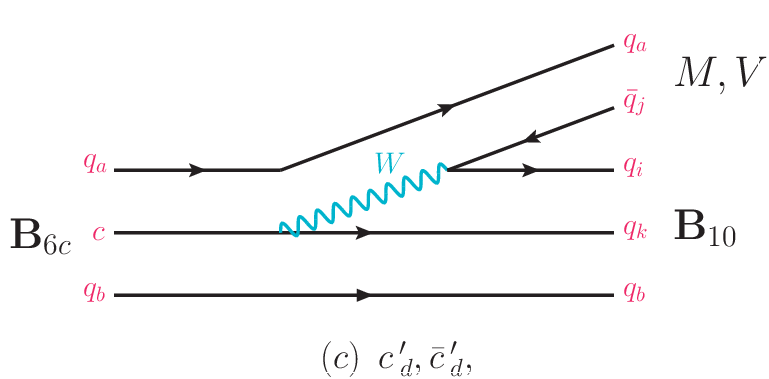}\\[2mm]
\includegraphics[width=1.8in]{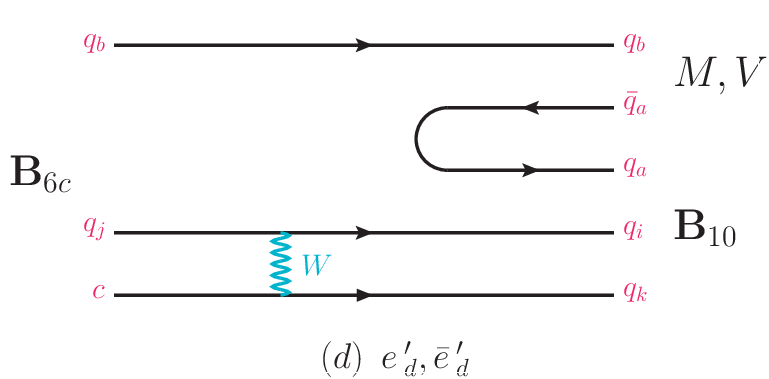}
\includegraphics[width=1.8in]{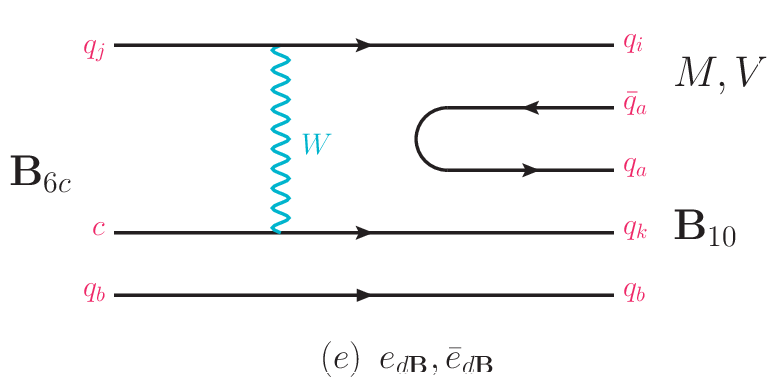}
\includegraphics[width=1.8in]{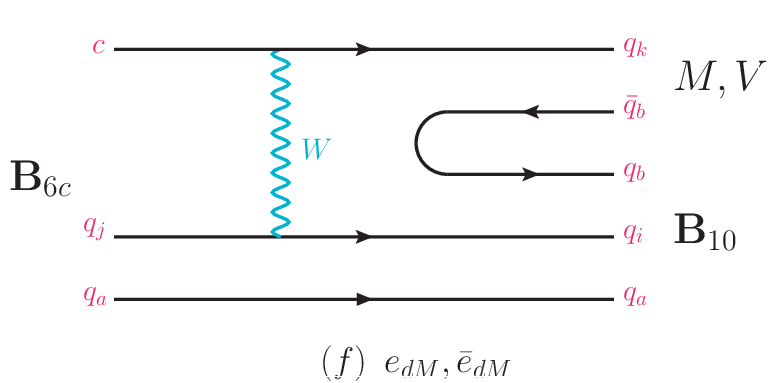}\\[2mm]
\caption{Topological diagrams of ${\bf B}_{6c}\to {\bf B}_{10}M(V)$ decays.}\label{fig2}
\end{figure}
%
We investigate potential two-body decays of $\Omega_c^0$ 
that occur through the Cabibbo-allowed (CA) and singly Cabibbo-suppressed (SCS)
charm quark weak transitions: $c\to su\bar d$ and $c\to qu\bar q$ with $q=(d,s)$, respectively.
The responsible effective Hamiltonian is defined as~\cite{Buchalla:1995vs,Buras:1998raa}
\begin{eqnarray}\label{Heff}
{\cal H}_{eff}&=&\sum_{i=+,-}\frac{G_F}{\sqrt 2}c_i
\left(V_{cs}^*V_{ud}O_i+V_{cq}^*V_{uq} O_i^q\right)\,.
\end{eqnarray}
Here, $G_F$ is the Fermi constant, $V_{ij}$ denotes 
the Cabibbo-Kobayashi-Maskawa (CKM) matrix element, and
$c_{\pm}$ are the scale-dependent Wilson coefficients accounting for
perturbative QCD corrections. At the scale $\mu=1.25$~GeV, one obtains 
$(c_+,c_-)=(0.71,1.98)$~\cite{Buchalla:1995vs,Zou:2019kzq}.
In Eq.~(\ref{Heff}), $O_\pm^{(q)}$ represent the four-quark operators,
which are defined as
\begin{eqnarray}\label{Opm}
&&
O_\pm={1\over 2}\left[(\bar u d)(\bar s c)\pm (\bar s d)(\bar u c)\right]\,,\;
\nonumber\\
&&
O_\pm^q={1\over 2}\left[(\bar u q)(\bar q c)\pm (\bar q q)(\bar u c)\right]\,.
\end{eqnarray}
In the expressions above, 
$(\bar q_1 q_2)\equiv\bar q_1\gamma_\mu(1-\gamma_5)q_2$.
We thus study the CA and SCS decay channels, whereas
the doubly Cabibbo-suppressed (DCS) two-body $\Omega_c^0$ decays 
are not within the scope of our investigation, as 
none of the DCS decay channels have been observed as those in the CA and CSC ones.

By omitting Lorentz indices,  
the effective Hamiltonian in two different $SU(3)_f$ representations
can be expressed as
${\cal H}_{eff}=(G_F/\sqrt 2) {\cal H}_{\rm TDA}$ and 
${\cal H}_{eff}=(G_F/\sqrt 2) {\cal H}_{\rm IRA}$.
In the context of the topological diagram approach, 
we consider flavor changes $c\to q_k q_i\bar q_j$ with $q_i=(u, d, s)$ 
represented as a triplet in the $SU(3)_f$ symmetry~\cite{Pan:2020qqo,Hsiao:2020iwc,Hsiao:2021nsc}:
\begin{eqnarray}\label{HTDA}
{\cal H}_{\rm{TDA}}=H_j^{ki}\,.
\end{eqnarray}
The non-zero entries in the equation are 
$H^{31}_2=1$ for $c\to s u\bar d$,
$H^{21}_2=-s_c$ for $c\to u d\bar d$, and
$H^{31}_3=s_c$ for $c\to u s\bar s$, where we have used
$V_{cs}^*V_{ud}\simeq 1$ and $V_{cd}^*V_{ud}=-V_{cs}^*V_{us}\simeq -s_c$
with the Cabbibo angle $\theta_c$ in $s_c\equiv \sin\theta_c$.

In the irreducible $SU(3)_f$ approach, the operators of Eq.~(\ref{Opm}) 
behave as $(\bar q^i q_k \bar q^j)c$ with respect to the $SU(3)_f$ symmetry, 
leading to $(\bar 3\times 3\times \bar 3)c=(\bar 3+\bar 3'+6+\overline{15})c$ 
in the irreducible form. Here, 6 and $\overline{15}$ correspond to 
$O_-\sim (\bar u d\bar s-\bar s d\bar u)c$ and $O_+\sim (\bar u d\bar s+\bar s d\bar u)c$
[$O^q_-\sim (\bar u q\bar q-\bar q q\bar u)c$ and $O^q_+\sim (\bar u q\bar q+\bar q q\bar u)c$],
respectively.
We thus obtain~\cite{Savage:1989qr,Savage:1991wu,Geng:2017esc}
\begin{eqnarray}\label{HIRA}
{\cal H}_{\text{IRA}}=c_- { \epsilon^{ijl} \over 2}H(6)_{lk}+c_+H(\overline{15})_{k}^{ij}\,.
\end{eqnarray}
The non-zero entries of $H(6)_{lk}$ are $H_{22}(6)=2$ and $H_{23,32}(6)=2s_c$, 
while those of $H(\overline{15})_k^{ij}$ are given by
$H_2^{13}(\overline{15})=H_2^{31}(\overline{15})=1$ and 
$H_2^{12,21}(\overline{15})=-H_3^{13,31}(\overline{15})=-s_c$~\cite{Savage:1989qr,Hsiao:2020iwc}.

We present $\Omega_c^0$ as a sextet charmed baryon: $({\bf B}_{6c})^3_3=\Omega_c^0$,
omitting other ${\bf B}_{6c}$ states that strongly decay.
The octet and decuplet baryons have components: 
\begin{eqnarray}\label{B8B10}
({\bf B}_8)^i_j&=&\left(\begin{array}{ccc}
\frac{1}{\sqrt{6}}\Lambda^0+\frac{1}{\sqrt{2}}\Sigma^0 & \Sigma^+ & p\\
 \Sigma^- &\frac{1}{\sqrt{6}}\Lambda^0 -\frac{1}{\sqrt{2}}\Sigma^0  & n\\
 \Xi^- & \Xi^0 &-\sqrt{\frac{2}{3}}\Lambda^0
\end{array}\right)\,,\nonumber\\
\sqrt{3}({\bf B}_{10})_{ijk}&=&
\left(\begin{array}{ccc}
\left(\begin{array}{ccc}
\sqrt{3}\Delta^{++}&\Delta^+ & \Sigma^{* +}\\
\Delta^+ &\Delta^0 & \frac{\Sigma^{* 0}}{\sqrt{2}}\\
\Sigma^{* +}& \frac{\Sigma^{* 0}}{\sqrt{2}}& \Xi^{* 0}
\end{array}\right),
\left(\begin{array}{ccc}
\Delta^+ &\Delta^0 & \frac{\Sigma^{* 0}}{\sqrt{2}}\\
\Delta^0 &\sqrt{3}\Delta^-& \Sigma^{* -}\\
\frac{\Sigma^{* 0}}{\sqrt{2}}&\Sigma^{* -}&\Xi^{* -}
\end{array}\right),
\left(\begin{array}{ccc}
\Sigma^{* +}& \frac{\Sigma^{* 0}}{\sqrt{2}}&\Xi^{* 0}\\
\frac{\Sigma^{* 0}}{\sqrt{2}}&\Sigma^{* -}&\Xi^{* -}\\
\Xi^{* 0}&\Xi^{* -}&\sqrt{3}\Omega^-
\end{array}\right)
\end{array}\right).
\end{eqnarray}
We also present the octet baryon as $({\bf B}_8)_{ijk} = \epsilon_{ijl} ({\bf B}_8)^l_k$.
As for another final state,
the usual octet pseudoscalar (vector) meson $M$($V$) has the following components:
\begin{eqnarray}
M^i_j&=&(\pi^{\pm,0},\;K^\pm,\;K^0,\;\bar K^0,\;\eta)\,,\nonumber\\
V^i_j&=&(\rho^{\pm,0},\;K^{*\pm},\;K^{*0},\;\bar K^{*0},\;\omega,\;\phi)\,.
\end{eqnarray}

Figures~\ref{fig1}$a$~and~\ref{fig1}$(b,c,d)$ 
represent the external and internal $W$-emission diagrams, respectively, and
Fig.~\ref{fig1}$(e,f,g,h,i,j)$ depicts the $W$-exchange diagrams.
Each $W$-exchange diagram requires an additional quark pair 
produced by a gluon from the vacuum, denoted by $g\to q\bar{q}$, 
where $q\bar{q}$ can be $u\bar{u}$, $d\bar{d}$, or $s\bar{s}$. This allows
$q$ and $\bar{q}$ to be distributed into ${\bf B}$ and $M$, respectively.
To parameterize the topological diagrams and derive the TDA amplitudes of $\Omega_c^0\to {\bf B} M$,
we connect ${\bf B}_{6c}$ to ${\cal H}_{\rm TDA}$, ${\bf B}_8$, and $M$ as follows:
\begin{eqnarray}\label{MTDA1}
&&{\cal M}_{\rm TDA}(\Omega_c^0\to {\bf B} M)\nonumber\\
&&=
t({\bf B}_{6c})^{ab} H^{ki}_j({\bf B}_8)_{kab}(M)_j^i+
c({\bf B}_{6c})^{ab} H^{ki}_j({\bf B}_8)_{iab}(M)_j^k\,\nonumber\\
&&+
c^\prime({\bf B}_{6c})^{ab} H^{ki}_j({\bf B}_8)_{ikb}(M)_j^a+
c^{\prime\prime}({\bf B}_{6c})^{ab} H^{ki}_j({\bf B}_8)_{ibk}(M)_j^a\,\nonumber\\
&&+
e^{\prime}({\bf B}_{6c})^{jb} H^{ki}_j({\bf B}_8)_{ika}(M)_a^b+
e^{\prime\prime}({\bf B}_{6c})^{jb} H^{ki}_j({\bf B}_8)_{iak}(M)_a^b\,\nonumber\\
&&+
e_{\bf B} ({\bf B}_{6c})^{jb} H^{ki}_j({\bf B}_8)_{kab}(M)_a^i+
e'_{\bf B}({\bf B}_{6c})^{jb} H^{ki}_j({\bf B}_8)_{kba}(M)_a^i\,\nonumber\\
&&+
e_M({\bf B}_{6c})^{jb} H^{ki}_j({\bf B}_8)_{iba}(M)_a^k+
e_M^\prime({\bf B}_{6c})^{jb} H^{ki}_j({\bf B}_8)_{iab}(M)_a^k\,.
\end{eqnarray}
In this equation, the decay topological diagrams 
in Figs.~\ref{fig1}a, \ref{fig1}b, \ref{fig1}c(d), \ref{fig1}e(f), \ref{fig1}g(h), and \ref{fig1}i(j)
are parameterized as $t$, $c$, $c^{\prime(\prime\prime)}$, $e^{\prime(\prime\prime)}$,
$e_{\bf B}^{(\prime)}$, and $e_M^{(\prime)}$, respectively. 
Specifically, $e_{\bf B}$ and $e_M$ correspond to the $W$-exchange diagrams
in Figs.~\ref{fig1}g and \ref{fig1}i, where the charm quark transforms into ${\bf B}$ and $M$, 
respectively. In contrast, $e^\prime$ represents the $W$-exchange diagram in Fig.~\ref{fig1}e,
where the $W$-boson exchange exclusively occurs within ${\bf B}$ without involving $M$.
Furthermore, $c^{\prime\prime}$, $e^{\prime\prime}$ and $e'_{{\bf B}(M)}$ 
parameterize the same $W$-exchange diagrams as $c'$,  $e'$ and $e_{{\bf B}(M)}$, respectively, 
but with a different anti-symmetric quark pair in ${\bf B}$. 

The decuplet baryon consists of quark contents that are totally symmetric, 
resulting in six topological diagrams for $\Omega_c^0\to{\bf B}^* M$ as drawn in Fig.~\ref{fig2}. 
The amplitudes are then derived as
\begin{eqnarray}\label{MTDA2}
&&{\cal M}_{\rm TDA}(\Omega_c^0\to{\bf B}^* M)\nonumber\\
&&=
t_d({\bf B}_{6c})^{ab} H^{ki}_j({\bf B}_{10})_{kab}(M)_j^i+
c_d({\bf B}_{6c})^{ab} H^{ki}_j({\bf B}_{10})_{iab}(M)_j^k\,\nonumber\\
&&+
c_d^{\prime}({\bf B}_{6c})^{ab} H^{ki}_j({\bf B}_{10})_{ikb}(M)_j^a+
e_d^{\prime}({\bf B}_{6c})^{jb} H^{ki}_j({\bf B}_{10})_{ika}(M)_a^b\,\nonumber\\
&&+
e_{d{\bf B}} ({\bf B}_{6c})^{jb} H^{ki}_j({\bf B}_{10})_{kab}(M)_a^i
+e_{dM}({\bf B}_{6c})^{jb} H^{ki}_j({\bf B}_{10})_{iba}(M)_a^k\,.
\end{eqnarray}
Here, $(t_d,c_d,c'_d)$ and $(e'_d, e_{d{\bf B}},e_{dM})$ parameterize
the $W$-emission and $W$-exchange topological diagrams 
of Fig.~\ref{fig2}$(a,b,c)$ and Fig.~\ref{fig2}$(d,e,f)$, respectively.

Similar to TDA, 
we connect ${\bf B}_{6c}$ to ${\cal H}_{\rm IRA}$, ${\bf B}_{8(10)}$, 
and $M$ to establish the IRA amplitudes of $\Omega_c^0\to {\bf B}^{(*)}M$. 
For $\Omega_c^0\to{\bf B}M$, the amplitudes are given by
\begin{eqnarray}\label{MIRA1}
&&{\cal M}_{\rm IRA}(\Omega_c^0\to {\bf B}M)=
{\cal M}_{6}+{\cal M}_{\overline{15}}\,,
\end{eqnarray}
where~\cite{Savage:1989qr,Geng:2017mxn}
\begin{eqnarray}
&&{\cal M}_{6}=
a_{12}H_{ij}(6)({\bf B}_{6c})^{ij}({\bf B}_8)_k^l(M)_l^k +
a_{13}H_{ij}(6)({\bf B}_{6c})^{kl}({\bf B}_8)_k^i(M)_l^j\nonumber\\
&&+
a_{14}H_{ij}(6)({\bf B}_{6c})^{jk}({\bf B}_8)_k^l(M)_l^i+
a_{15}H_{ij}(6)({\bf B}_{6c})^{jk}({\bf B}_8)_l^i(M)_k^l\,,\nonumber\\
&&{\cal M}_{\overline{15}}=
a_{16}({\bf B}_8)_j^i(M)_l^k H(\overline{15})^{jm}_i({\bf B}_{6c})^{ln}\epsilon_{kmn}+
a_{17}({\bf B}_8)_j^i(M)_l^k H(\overline{15})^{lm}_i({\bf B}_{6c})^{jn}\epsilon_{kmn}\nonumber\\
&&+
a_{18}({\bf B}_8)_n^m(M)_j^n H(\overline{15})^{ij}_k({\bf B}_{6c})^{kl}\epsilon_{ilm}+
a_{19}({\bf B}_8)_l^j(M)_n^k H(\overline{15})^{il}_m({\bf B}_{6c})^{mn}\epsilon_{ijk}\nonumber\\
&&+
a_{20}({\bf B})_n^j(M)_l^k H(\overline{15})^{il}_m({\bf B}_{6c})^{mn}\epsilon_{ijk}\,.
\end{eqnarray}
The parameters $a_i$ (with $i$ ranging from 12 to 20) represent the $SU(3)_f$ invariant amplitudes.
For $\Omega_c^0\to{\bf B}^*M$, we express the amplitudes to be~\cite{Savage:1989qr,Geng:2017mxn}
\begin{eqnarray}\label{MIRA2}
&&
{\cal M}_{\rm IRA}(\Omega_c^0\to{\bf B}^* M)={\cal M}'_{6}+{\cal M}'_{\overline{15}}\,,
\end{eqnarray}
where ${\cal M}'_6$ and ${\cal M}'_{\overline{15}}$ 
are defined similarly to ${\cal M}_6$ and ${\cal M}_{\overline{15}}$, respectively,
given by
\begin{eqnarray}
&&
{\cal M}'_{6}=
a_{21}({\bf B}_{10})_{lkm}(M)_n^i H_{ij}(6)({\bf B}_{6c})^{lk}\epsilon^{jmn} +
a_{22}({\bf B}_{10})_{klm}(M)_n^l H_{ij}(6)({\bf B}_{6c})^{jk}\epsilon^{imn} \,,\nonumber\\
&&{\cal M}'_{\overline{15}}=
a_{23}({\bf B}_{10})_{ijk}(M)_l^m  H(\overline{15})^{lk}_m ({\bf B}_{6c})^ {ij}+
a_{24}({\bf B}_{10})_{ijk}(M)_m^k  H(\overline{15})^{ij}_l ({\bf B}_{6c})^{lm}\nonumber\\
&&+
a_{25}({\bf B}_{10})_{ijk}(M)_m^l  H(\overline{15})^{ij}_l ({\bf B}_{6c})^{km}+
a_{26}({\bf B}_{10})_{ijk}(M)_l^j  H(\overline{15})^{lk}_m ({\bf B}_{6c})^{im}\,.
\end{eqnarray}
The parameters $a_i$ (with $i$ ranging from 21 to 26) are another set of the $SU(3)_f$ parameters.

Using ${\cal M}_{\rm TDA}(\Omega_c^0 \to {\bf B}^{(*)}M) = {\cal M}_{\rm IRA}(\Omega_c^0 \to{\bf B}^{(*)}M)$, 
where the expansions can be found in Table~\ref{tab1}, we unify the two $SU(3)_f$ approaches and 
establish the equivalence relations of the two-body $\Omega_c^0$ decays for the first time
\begin{eqnarray}\label{re1}
&&(t,c,c',c^{\prime\prime})=(-2a_{14}+a_{17},2a_{14}+a_{17},-2a_{13}-2a_{14},0)\,,\nonumber\\
&&(e',e^{\prime\prime})=(-2a_{12}-2a_{15}+a_{19},-2a_{19})\,,\nonumber\\
&&(e_{\bf B},e'_M)=(-a_{17}-a_{20},-a_{17}-a_{20})\,,\nonumber\\
&&(e_M,e'_{\bf B})=(2a_{12}+a_{18},-2a_{12}+a_{18})\,,
\end{eqnarray}
for $\Omega_c^0\to {\bf B}M$
and
\begin{eqnarray}\label{re2}
&&
(t_d,c_d,c'_d,e'_d)=
(2a_{21}+a_{23},-2a_{21}+a_{23},2a_{25},2a_{24})\,,\nonumber\\
&&
(e_{d\bf B},e_{dM})=(2a_{22}+a_{26},-2a_{22}+a_{26})\,,
\end{eqnarray}
for $\Omega_c^0 \to {\bf B}^*M$. Therefore,
the two $SU(3)_f$ parameters  
are equivalent in $\Omega_c^0 \to {\bf B}^{(*)}M$ as we have demonstrated in 
${\bf B}_{3c}\to {\bf B}^{(*)}M$~\cite{He:2018php,He:2018joe,Hsiao:2020iwc,Hsiao:2021nsc}.

The TDA helps to visualize the decay processes of $\Omega_c^0\to{\bf B}M,{\bf B}^*M$,
leading to a more intuitive understanding for the decays. Apart from this advantage, 
the current data as presented in Eq.~(\ref{data1}) are not sufficient for a global fit.
Fortunately, it is possible that its unification with IRA in Eqs.~(\ref{re1}), (\ref{re2}) can be used to
combine or neglect the topological parameters, thus reducing the parameters 
for a practical extraction. 

More explicitly, the reduction is based on the irreducible form of 
${\cal H}_{\text{IRA}}\propto c_- H(6)+c_+ H(\overline{15})$ in Eq.~(\ref{HIRA}), 
where $c_+/c_- \sim 35\%$ clearly favors the sextet part of ${\cal H}_{\rm IRA}$. 
Thus, the sextet-dominance relations have been proposed 
for charmed baryon decays~\cite{Savage:1989qr, Lu:2016ogy, Li:2012cfa}. 
As a result, the parameters $(a_{12}, a_{13}, a_{14}, a_{15})$ for $\Omega_c^0\to{\bf B}M$ 
and $(a_{21}, a_{22})$ for $\Omega_c^0\to{\bf B}^*M$ absorb $c_-$ and become dominant, 
while the parameters associated with $c_+ H(\overline{15})$ 
are expected to contribute around a 35\% correction to the amplitudes. 
This correction is roughly comparable to the $SU(3)_f$ symmetry breaking effect, 
estimated at around 20--30\% and can therefore be considered negligible. 
Consequently, the neglect of the 15-plet IRA parameters has been widely applied in the literature, 
especially when measurements were once insufficient for a statistically significant global fit.
Remarkably, this approach has proven effective 
in explaining and predicting branching fractions~\cite{Lu:2016ogy, Geng:2017mxn, 
Geng:2018bow, Geng:2017esc, Geng:2018plk, Geng:2018upx, Hsiao:2019yur}.

Utilizing the negligible characterization, 
we derive the following approximate equivalence relations from Eqs.~(\ref{re1}) and (\ref{re2}):
\begin{eqnarray}\label{re3}
&&
t\simeq -c \simeq -2a_{14}\,,c' \simeq -2a_{13}-2a_{14}\,,
e' \simeq -e_M \simeq e'_{\bf B} \simeq -2a_{12}\,,
(e^{\prime\prime}, e_{\bf B}, e'_M)\simeq 0\,, 
\end{eqnarray}
for $\Omega_c^0\to{\bf B}M$, with $c^{\prime\prime}=0$, and 
\begin{eqnarray}\label{re3b}
&&
t_d \simeq -c_d \simeq 2a_{21}\,,
e_{d\bf B} \simeq -e_{dM}\simeq 2a_{22}\,,
(c'_d, e'_d)\simeq 0\,,
\end{eqnarray}
for $\Omega_c^0\to {\bf B}^* M$.
Note that the parameter $e'_{\bf B}$ appears only in the DCS decay channels, 
such as ${\cal M}(\Omega_c^0 \to \Sigma^-\pi^+)$, 
which are not within the scope of the current study. 
However, $e'_{\bf B}$ is included in Eqs.~(\ref{re1}), (\ref{re3})
to complete the (approximate) equivalence relations.

In Eq.~(\ref{MTDA1}), $c^{\prime\prime}$ and $e^{\prime\prime}$ 
are included to complete our derivation of the amplitudes for $\Omega_c^0\to {\bf B}M$.
However, the $c^{\prime\prime}$ and $e^{\prime\prime}$-like terms are absent
in the derivation of the amplitudes for ${\bf B}_{3c}\to {\bf B}^{(*)}M$ in Ref.~\cite{Kohara:1991ug}.
This absence is grounded in the K$\ddot{\text{o}}$rner-Pati-Woo (KPW)
theorem~\cite{Miura:1967lka,Korner:1970xq,Pati:1970fg}, 
which states that in weak currents, such as $(\bar q_i q_j)(\bar q_k c)$, 
the quarks $q_i$ and $q_k$, when connected by an exchange of a $W$-boson, 
must exhibit color anti-symmetry.
This results in their flavor anti-symmetry as the constituents of the octet baryon. 
As illustrated in Fig.~\ref{fig1}, using $c'$ and $e'$, 
rather than $c^{\prime\prime}$ and $e^{\prime\prime}$, adheres to this theorem.
On the other hand, we find $c^{\prime\prime}=0$ in Eq.~(\ref{re1}) 
and $e^{\prime\prime}\simeq 0$ in Eq.~(\ref{re3}). Therefore,
in one way or another, 
$c^{\prime\prime}$ and $e^{\prime\prime}$ should be neglected. 

By adding the bar notation to the parameters and replacing 
$(M)^i_j$ with $(V)^i_j$ in Eqs.~(\ref{MTDA1}), (\ref{MTDA2}) and Eqs.~(\ref{MIRA1}), (\ref{MIRA2}), 
we get ${\cal M}_{\rm TDA,IRA}(\Omega_c^0\to {\bf B}^{(*)}V)$ as presented in Table~\ref{tab1}. 
This results in equivalence relations the same as those in Eqs.~(\ref{re1}), (\ref{re2}) and (\ref{re3}),
except for the added bar notation. Consequently, we simplify the TDA amplitudes as
${\cal M}_{\rm STDA}(\Omega_c^0\to{\bf B}^{(*)}M,{\bf B}^{(*)}V)$ 
using the approximate equivalence relations in Eq.~(\ref{re3}), as listed in Table~\ref{tab3}.
%
\begin{table}[b]
\caption{${\cal M}_{\rm{TDA,IRA}}(\Omega_c^0\to{\bf B}M(V),{\bf B}^*M(V))$
from the expansions in Eqs.~(\ref{MTDA1}, \ref{MTDA2})~and~(\ref{MIRA1}, \ref{MIRA2}),
respectively.}\label{tab1}
\tiny
\begin{tabular}{|l|l|l|}
\hline
Decay mode
&$\;\;\;\;\;\;\;\;\;\;\;\;\;\;\;\;\;\;\;\;\;\;\;\;\;\;\;$ ${\cal M}_{\text{TDA}}$
&$\;\;\;\;\;\;\;\;\;\;\;\;\;\;\;\;\;\;\;\;\;\;\;\;\;\;\;\;\;\;\;\;\;\;\;\;\;\;\;\;\;\;\;\;$ ${\cal M}_{\text{IRA}}$ \\
\hline\hline
$\Omega_c^0 \to \Xi^0 \bar{K}^{0}$
&$-(c+c^\prime)$
&$2a_{13}-a_{16}-a_{17}$
\\
$\Omega_c^0 \to \Lambda^0 \bar{K}^{0}$
&$\frac{1}{\sqrt{6}}(2c^\prime+c^{\prime \prime}-e_M-2e_M^\prime-e^{\prime}-2e^{\prime \prime})s_c$
&$\frac{-1}{\sqrt{6}}(4a_{13}+4a_{14}-2a_{15}-3a_{16}-2a_{17}+a_{18}-3a_{19}-2a_{20})s_c$\\
$\Omega_c^0 \to \Sigma^0 \bar{K}^{0}$
&$\frac{1}{\sqrt{2}}(e_M+e^\prime-c^{\prime \prime})s_c$
&$ \frac{-1}{\sqrt{2}}(2a_{15}+a_{16}-a_{18}-a_{19})s_c$
\\
$\Omega_c^0 \to \Sigma^+ K^{-}$
&$-(e_{{M}}+e^\prime)s_c$
&$(2a_{15}-a_{16}+a_{18}+a_{19})s_c$
\\
$\Omega_c^0 \to \Xi^- \pi^+$
&$-(t+e_{{\bf B}})s_c$
&$(2a_{14}+a_{20})s_c $ \\
$\Omega_c^0 \to \Xi^0 \pi^0$
&$\frac{-1}{\sqrt{2}}(c-e_{{\bf B}})s_c$
&$\frac{-1}{\sqrt{2}}(2a_{14}+2a_{17}+a_{20})s_c$
\\
$\Omega_c^0\to \Xi^0 \eta$
&$\frac{1}{\sqrt{6}}(3c+2c^\prime+2c^{\prime \prime}+e_{\bf{B}}+2e_M^{(s)}+2e_M^{\prime (s)}+2e^{\prime(s)}+2e^{\prime \prime(s)})s_c$
&$\frac{-1}{\sqrt{6}}(4a_{13}-2a_{14}+4a_{15}-2a_{18}+2a_{19}+3a_{20})s_c$\\
\hline
$\Omega_c^0 \to \Xi^0 \bar{K}^{*0}$
&$-(\bar{c}+\bar{c}^\prime)$
&$2\bar{a}_{13}-\bar{a}_{16}-\bar{a}_{17}$
\\
$\Omega_c^0 \to \Lambda^0 \bar{K}^{*0}$
&$\frac{1}{\sqrt{6}}(2\bar{c}^\prime+\bar{c}^{\prime \prime}-\bar{e}_M
-2\bar{e}_M^\prime-\bar{e}^{\prime}-2\bar{e}^{\prime \prime})s_c$
&$\frac{-1}{\sqrt{6}}(4\bar{a}_{13}+4\bar{a}_{14}-2\bar{a}_{15}-3\bar{a}_{16}
-2\bar{a}_{17}+\bar{a}_{18}-3\bar{a}_{19}-2\bar{a}_{20})s_c$\\
$\Omega_c^0 \to \Sigma^0 \bar{K}^{*0}$
&$\frac{1}{\sqrt{2}}(\bar{e}_M+\bar{e}^\prime-\bar{c}^{\prime \prime})s_c$
&$ \frac{-1}{\sqrt{2}}(2\bar{a}_{15}+\bar{a}_{16}-\bar{a}_{18}-\bar{a}_{19})s_c$
\\
$\Omega_c^0 \to \Sigma^+ K^{*-}$
&$-(\bar{e}_{{M}}+\bar{e}^\prime)s_c$
&$(2\bar{a}_{15}-\bar{a}_{16}+\bar{a}_{18}+\bar{a}_{19})s_c$
\\
$\Omega_c^0 \to \Xi^- \rho^+$
&$-(\bar{t}+\bar{e}_{{\bf B}})s_c$
&$(2\bar{a}_{14}+\bar{a}_{20})s_c $ \\
$\Omega_c^0 \to \Xi^0 \rho^0$
&$\frac{-1}{\sqrt{2}}(\bar{c}-\bar{e}_{{\bf B}})s_c$
&$\frac{-1}{\sqrt{2}}(2\bar{a}_{14}+2\bar{a}_{17}+\bar{a}_{20})s_c$
\\
$\Omega_c^0\to \Xi^0 \omega$
&$\frac{1}{\sqrt{2}}(\bar{c}+\bar{e}_{\bf{B}})s_c$
&$\frac{1}{\sqrt{2}}(2\bar{a}_{14}-\bar{a}_{20})s_c$\\
$\Omega_c^0\to \Xi^0 \phi$
&$-(\bar{c}+\bar{c}^\prime+\bar{c}^{\prime \prime}+\bar{e}_M^{(s)}+\bar{e}_M^{\prime(s)}+\bar{e}^{\prime(s)}+\bar{e}^{\prime\prime(s)})s_c$
&$(2\bar{a}_{13}+2\bar{a}_{15}-\bar{a}_{18}+\bar{a}_{19}+\bar{a}_{20})s_c$\\
\hline
\hline
$\Omega_c^0 \to \Xi^{* 0} \bar{K}^{0}$
&$\frac{1}{\sqrt 3}(c_d+c^\prime_d)$
&$\frac{-1}{\sqrt 3}(2a_{21}-a_{23}-2a_{25})$\\
$\Omega_c^0 \to \Omega^- \pi^+$
&$t_d$
&$2a_{21}+a_{23}$
\\
$\Omega_c^0 \to \Omega ^- K^{+}$
&$(t_d+e_{d{\bf B}}^{(s)})s_c$
&$(2a_{21}+2a_{22}+a_{23}+a_{26})s_c$
\\
$\Omega_c^0 \to \Sigma^{* +}K^{-}$
&$\frac{1}{\sqrt 3}(e_{dM}+e^\prime_d)s_c$
&$\frac{-1}{\sqrt 3}(2a_{22}-2a_{24}-a_{26})s_c$
\\
$\Omega_c^0 \to \Sigma^{*0} \bar K^{0}$
&$\frac{-1}{\sqrt 6}(c^\prime_d-e_{dM}-e^\prime_d)s_c$
&$\frac{-1}{\sqrt 6}(2a_{22}-2a_{24}+2a_{25}-a_{26})s_c$
\\
$\Omega_c^0 \to \Xi^{*-} \pi^+$
&$\frac{-1}{\sqrt 3}(t_d-e_{d{\bf B}})s_c$
&$\frac{-1}{\sqrt 3}(2a_{21}-2a_{22}+a_{23}-a_{26})s_c $
\\
$\Omega_c^0 \to \Xi^{*0} \pi^0$
&$\frac{1}{\sqrt 6}(c_d+e_{d{\bf B}})s_c$
&$\frac{-1}{\sqrt 6}(2a_{21}-2a_{22}-a_{23}-a_{26})s_c$
\\
$\Omega_c^0\to \Xi^{* 0} \eta$
&$\frac{-1}{\sqrt 18}(3c_d+2c^\prime_d-e_{d{\bf B}}+2e_{dM}^{(s)}+2e^{\prime(s)}_d)s_c$
&$\frac{-1}{\sqrt 18}(6a_{21}+6a_{22}-3a_{23}-24a_{24}-4a_{25}-a_{26})s_c$
\\
\hline
$\Omega_c^0 \to \Xi^{* 0} \bar{K}^{*0}$
&$\frac{1}{\sqrt 3}(\bar{c}_d+\bar{c}^\prime_d)$
&$\frac{-1}{\sqrt 3}(2\bar{a}_{21}-\bar{a}_{23}-2\bar{a}_{25})$\\
$\Omega_c^0 \to \Omega^-\rho^+$
&$\bar{t}_d$
&$2\bar{a}_{21}+\bar{a}_{23}$
\\
$\Omega_c^0 \to \Omega^- K^{*+}$
&$(\bar{t}_d+\bar{e}_{d{\bf B}}^{(s)})s_c$
&$(2\bar{a}_{21}+2\bar{a}_{22}+\bar{a}_{23}+\bar{a}_{26})s_c$
\\
$\Omega_c^0 \to \Sigma^{*+}K^{*-}$
&$\frac{1}{\sqrt 3}(\bar{e}_{dM}+\bar{e}^\prime_d)s_c$
&$\frac{-1}{\sqrt 3}(2\bar{a}_{22}-2\bar{a}_{24}-\bar{a}_{26})s_c$
\\
$\Omega_c^0 \to \Sigma^{*0} \bar K^{*0}$
&$\frac{-1}{\sqrt 6}(\bar{c}^\prime_d-\bar{e}_{dM}-\bar{e}^\prime_d)s_c$
&$\frac{-1}{\sqrt 6}(2\bar{a}_{22}-2\bar{a}_{24}+2\bar{a}_{25}-\bar{a}_{26})s_c$
\\
$\Omega_c^0 \to \Xi^{*-} \rho^+$
&$\frac{-1}{\sqrt 3}(\bar{t}_d-\bar{e}_{d{\bf B}})s_c$
&$\frac{-1}{\sqrt 3}(2\bar{a}_{21}-2\bar{a}_{22}+\bar{a}_{23}-\bar{a}_{26})s_c $
\\
$\Omega_c^0 \to \Xi^{*0} \rho^0$
&$\frac{1}{\sqrt 6}(\bar{c}_d+\bar{e}_{d{\bf B}})s_c$
&$\frac{-1}{\sqrt 6}(2\bar{a}_{21}-2\bar{a}_{22}-\bar{a}_{23}-\bar{a}_{26})s_c$
\\
$\Omega_c^0 \to \Xi^{*0} \omega$
&$\frac{-1}{\sqrt 6}(\bar{c}_d-\bar{e}_{d{\bf B}})s_c$
&$\frac{1}{\sqrt 6}(2\bar{a}_{21}+2\bar{a}_{22}-\bar{a}_{23}+\bar{a}_{26})s_c$
\\
$\Omega_c^0 \to \Xi^{*0} \phi$
&$\frac{1}{\sqrt 3}(\bar{c}_d+\bar{c}^\prime_d+\bar{e}_{dM}^{(s)}+\bar{e}^{\prime(s)}_d)s_c$
&$\frac{-1}{\sqrt 3}(2\bar{a}_{21}+2\bar{a}_{22}-\bar{a}_{23}-2\bar{a}_{24}-2\bar{a}_{25}-\bar{a}_{26})s_c$
\\
\hline
\end{tabular}
\end{table}
%

Under the $SU(3)_f$ symmetry, $m_s\simeq m_{u,d}$ is assumed. However,
in reality, $m_s$ is larger than $m_{u,d}$, which introduces possible $SU(3)_f$ breaking effects.
Following~Ref.~\cite{Gronau:1995hm}, we present the sources of the breaking effects 
term-by-term within the topological-diagram scheme
by analyzing the $s$-quark flavor flows as outlined in:
(i)~spectator $s$-quark(s) within the ${\bf B}_{3c,6c}$ baryon,
(ii)~weak transitions $c\to su\bar d$ and $c\to u s\bar s$, and
(iii)~$g\to s\bar s$ in $W$-exchange diagrams.

We use the commonly applied parameterization to estimate the breaking effects 
in meson production~\cite{Berthiaume:2023kmp,He:2000ys}. 
Due to the $s$-quark flavor flows from points (i) and (ii), 
we have $\delta_M=(f_K-f_\pi)/f_\pi$, $\delta_{V_1}=(f_{K^*}-f_\rho)/f_\rho$, 
and $\delta_{V_2}=(f_{\phi}-f_{\omega})/f_\omega$, 
where $f_{M(V)}$ represents the decay constant~\cite{Hsiao:2014mua,Hsiao:2017tif}.
Similarly, for baryon production, the breaking effect is estimated as 
$\delta_{\bf B}=(F_{\Omega}-F_{\Xi^*})/F_{\Xi^*}$~\cite{He:2000ys,Li:2012cfa,Cheng:2019ggx}, 
where $F_{\bf B}$ refers to 
the $\Omega_c^0 \to \Omega(\Xi^*)$ transition form factor~\cite{Hsiao:2021mlp,Hsiao2025}. 
Additionally, $g \to s\bar s$ in (iii) pertains specifically to non-factorizable $W$-exchange diagrams. 
We denote the $W$-exchange parameters involving $g \to s\bar{s}$ with the subscript $(s)$ in Table~\ref{tab1}, 
such that the $g \to s\bar s$-induced breaking effects are estimated as $\delta_E=(e^s-e)/e$.

While applying the simplified TDA (STDA), 
it remains unclear if any discrepancies with the data stem from $SU(3)_f$-breaking effects 
or from the 15-plet parameters excluded in the sextet-dominance relations, 
as both types of contributions generally have similar magnitudes in adjusting  
amplitudes or branching fractions. To account for these corrections simultaneously, 
we use $\delta_c=c_+/c_-\simeq 35\%$ in our estimates.
Consequently, the possible $SU(3)_f$ symmetry breaking results in 
$\delta_M\simeq 0.6\delta_c$, $\delta_{V_1,V_2}\simeq (0.2,0.4)\delta_c$, and $\delta_{\bf B}\simeq 0.3\delta_c$. 
Although $\delta_E$ cannot be theoretically estimated, 
the ratios ${\cal B}(D^0\to K^+ K^-)/{\cal B}(D^0\to \pi^+\pi^-)$~\cite{Li:2012cfa,Cheng:2019ggx}, 
${\cal B}(\Lambda_c^+ \to \Sigma^{*+} \eta)$~\cite{Hsiao:2020iwc}, 
and ${\cal B}(\Xi_c^0\to \Xi^- K^+)/{\cal B}(\Xi_c^0\to \Xi^- \pi^+)$~\cite{Hsiao:2021nsc} 
suggest that $\delta_E$ could be approximately 40--70\%, roughly $(1-2)\times \delta_c$.

Following~\cite{Gronau:1995hm}, we consider the $t$-like term as the dominant contribution 
and estimate corrections to the topological parameters relative to the $t$-like term. 
These corrections account for the neglected $c_+$ terms and possible $SU(3)_f$ breaking effects.
The order-by-order estimates in powers of $\delta_c$ are presented as follows:\\
(i)~${\cal O}(\delta_c^0)$: 
the dominant contributions to ${\cal M}(\Omega_c^0\to{\bf B}M)$ and 
${\cal M}(\Omega_c^0\to{\bf B}^* M)$ are given by $t$ and $t_d$, respectively.\\
(ii)~${\cal O}(\delta_c^1)$: 
contributions from $(e^{\prime\prime(s)}, e_{\bf B}, e_M^{\prime (s)})$
and corrections due to the neglected $c_+$ terms 
to $(c, c',e',e_M^{(s)},e'_{\bf B})$ for $\Omega_c^0\to{\bf B}M$;
contributions from $(c'_d, e_d^{\prime (s)})$ and 
corrections due to the neglected $c_+$ terms to $(t_d,c_d,e_{d\bf B}^{(s)},e_{dM}^{(s)})$
for $\Omega_c^0\to{\bf B}^* M$. In particular, 
$\delta_M$ and $\delta_E$  from $SU(3)_f$-breaking effects 
are anticipated to be of order $\delta_c$, 
whereas $\delta_{V_1,V_2}$ and $\delta_{\bf B}$ 
contribute minimally.\\
(iii)~${\cal O}(\delta_c^2)$: $SU(3)_f$-breaking effects for 
$(e^{\prime\prime(s)},e_M^{\prime s})$ and $e_d^{\prime s}$.\\
(iv)~These order-by-order estimates also extend to the $\Omega_c^0\to {\bf B}^{(*)}V$ cases.

The $SU(3)_f$-breaking effects have been scarcely observed in charmed baryon decays. 
As shown in Ref.~\cite{Xing:2023dni}, the $\chi^2$ value relative to the degrees of freedom 
indicates a reasonable fit under exact $SU(3)_f$ symmetry. 
This aligns with our order-by-order estimates, 
where $\delta_E \simeq (1-2)\times \delta_c$ suggests that 
only the exchange diagrams are likely to produce larger symmetry-breaking effects 
in experimental measurements. As we will discuss later, 
some specific decays, such as ${\cal B}(\Lambda_c^+ \to \Sigma^{*+} \eta)$~\cite{Hsiao:2020iwc} 
and ${\cal B}(\Xi_c^0\to \Xi^- K^+)/{\cal B}(\Xi_c^0\to \Xi^- \pi^+)$~\cite{Hsiao:2021nsc}, 
may hint at $SU(3)_f$ flavor breaking induced by the $g\to s\bar s$ transition.

To determine the topological parameters,
we need an equation to turn the amplitude into the branching fraction, 
given by~\cite{pdg}
\begin{eqnarray}\label{p_space}
&&{\cal B}(\Omega_c^0\to{\bf B}^{(*)}M,{\bf B}^{(*)}V)=
\frac{G_F^2|\vec{p}_{\rm cm}|\tau_{\Omega_c}}{16\pi m_{\Omega_c}^2}
|{\cal M}_{\rm STDA}(\Omega_c^0\to{\bf B}^{(*)}M,{\bf B}^{(*)}V)|^2\,,
\end{eqnarray}
with $|\vec{p}_{\rm cm}|
=[(m_{\Omega_c}^2-m_+^2)(m_{\Omega_c}^2-m_-^2)]^{1/2}/(2 m_{\Omega_c})$,
where $m_\pm=m_{{\bf B}^{(*)}}\pm m_{M(V)}$, 
$\tau_{\Omega_c}$ stands for the lifetime of $\Omega_c^0$, and
$\vec{p}_{cm}$ is the three-momentum of the final state in the $\Omega_c^0$ rest frame.

\section{Numerical Analysis}
To perform a numerical analysis, we use the Wolfenstein parameter $\lambda=s_c=0.225$
to present the CKM matrix elements as~\cite{pdg}
\begin{eqnarray}\label{CKM}
&&(V_{cs},V_{ud},V_{us},V_{cd})=(1-\lambda^2/2,1-\lambda^2/2,\lambda,-\lambda)\,.
\end{eqnarray}
Using one theoretical branching fraction as an input
for the experimental rates in Eq.~(\ref{data1}),
we can extract additional information.
As for the candidates, the branching fractions of $\Omega_c^0\to \Omega^- e^+\nu_e$, 
$\Omega^- \pi^+$, and $\Omega^- \rho^+$ have been calculated with 
the $\Omega_c^0\to\Omega^-$ transition form factors 
studied in~\cite{Pervin:2006ie,Hsiao:2020gtc,Xu:1992sw,Cheng:1996cs,Gutsche:2018utw}.
Since the semileptonic decay involves a lepton pair free 
from QCD corrections~\cite{Hsiao:2023qtk,Ke:2023qzc}, 
${\cal B}_e\equiv{\cal B}(\Omega_c^0\to \Omega^- e^+\nu_e)$ 
can be considered more reliable than 
${\cal B}(\Omega_c^0\to \Omega^-\pi^+,\Omega^- \rho^+)$. 
Therefore, we follow Ref.~\cite{Hsiao:2020gtc} to calculate 
${\cal B}_e=(5.4\pm 0.2)\times 10^{-3}$. 
By relating ${\cal B}_e$ to ${\cal R}_e$ in Eq.~(\ref{data1}), we extract 
${\cal B}_\pi\equiv{\cal B}(\Omega_c^0\to\Omega^-\pi^+) = (29.6\pm2.5)\times 10^{-4}$,
instead of calculating it. We then use ${\cal B}_\pi$ 
to extract other absolute branching ratios, as given in Table~\ref{tab3}.

The study of the branching fractions relies on
the determination of the topological parameters of ${\cal M}_{\rm STDA}$ in Table~\ref{tab3},
that is, $(t,c')$ for $\Omega_c^0\to {\bf B}M$, $(\bar t,\bar c')$ for $\Omega_c^0\to {\bf B}V$,
$(t_d,e_{dM})$ for $\Omega_c^0\to {\bf B}^*M$, and 
$(\bar t_d,\bar e_{dM})$ for $\Omega_c^0\to {\bf B}^*V$, 
where the approximate equivalence relations in Eqs.~(\ref{re3}), (\ref{re3b}), such as
$c_{(d)}\simeq -t_{(d)}$ and $\bar c_{(d)}\simeq -\bar t_{(d)}$, have been applied.
We can use the absolute branching fractions in Table~\ref{tab3} to extract the topological parameters.
By utilizing ${\cal B}(\Omega_c^0 \to \Xi^0 \bar{K}^{0},\Xi^-\pi^+)=(48.5\pm9.5,4.8\pm0.5)\times 10^{-4}$ and  
${\cal B}(\Omega_c^0 \to \Omega^- \pi^+,\Omega ^- K^{+})=(29.6\pm2.5,1.8\pm0.3)\times 10^{-4}$
in Table~\ref{tab3} as the data inputs, associated with 
${\cal M}_{\rm STDA}(\Omega_c^0 \to \Xi^0 \bar{K}^{0},\Xi^-\pi^+)$ and
${\cal M}_{\rm STDA}(\Omega_c^0 \to \Omega^- \pi^+,\Omega ^- K^{+})$, 
respectively, we can determine the values of $(t,c')$ and $(t_d,e_{dM})$.

For $\Omega_c^0\to{\bf B}V$, the single data point in Table~\ref{tab3}
is insufficient to determine $\bar t$ and $\bar c'$ simultaneously. 
Understanding that terms represented by $t$ and $\bar t$ 
correspond to factorizable amplitudes~\cite{Hsiao:2020gtc,Hsiao:2021mlp}, expressed as:
\begin{eqnarray}\label{Mfac}
{\cal M}_{\rm fac}(\Omega_c^0 \to {\bf B}M,{\bf B}V)
=\frac{G_F}{\sqrt 2}V_{cq}^* V_{q_1 q_2} a_1
\langle M,V|(\bar q_1q_2)|0\rangle\langle{\bf B}|(\bar q c)|\Omega_c^0\rangle\,,
\end{eqnarray}
can aid in practical determinations.
Utilizing Eq.~(\ref{Mfac}) and the $\Omega_c^0\to\Xi^-$ transition form factors in~\cite{Hu:2020nkg},
we calculate the branching fraction of $\Omega_c^0 \to \Xi^- \pi^+$.
Connecting our calculation to the data input in Table~\ref{tab3},
${\cal B}(\Omega_c^0 \to \Xi^- \pi^+)=(4.8\pm0.5)\times 10^{-4}$,
we extract $a_1=0.94\pm 0.05$. 
Given that $a_1$ is of order 1.0, consistent with the empirical input
from the generalized factorization~\cite{Bauer:1986bm,Ali:1998eb,
Hsiao:2019ann,Hsiao:2022tfj,Hsiao:2021mlp,Hsiao:2019wyd,Hsiao:2020gtc}, 
the feasibility of our data collection and extraction for $\Omega_c^0\to\Xi^- \pi^+$
can be confirmed.

We then use $a_1$ as an input to obtain the numerical result:
${\cal B}(\Omega_c^0 \to \Xi^- \rho^+)=(16.6\pm1.2)\times 10^{-4}$. 
In addition to ${\cal B}(\Omega_c^0 \to \Xi^0 \bar{K}^{*0})=(30.2\pm7.5)\times 10^{-4}$ 
in Table~\ref{tab3}, we have two inputs to get $\bar t$ and $\bar c'$ simultaneously.
With ${\cal B}(\Omega_c^0 \to \Xi^{*0} \bar{K}^{*0},\Omega^-\rho^+)
=(15.1\pm5.2,53.3\pm10.8)\times 10^{-4}$, we can fit $\bar t_d$ for $\Omega_c^0\to {\bf B}^* V$,
whereas the branching fraction for $\bar e_{dM}$ has not yet been measured. 
For estimation, 
we set $\bar e_{dM}=(e_{dM}/t_d)\bar t_d$.

Since the non-factorizable parameters can destructively or constructively interfere
with the $t$-like terms, there can be two solutions for our extraction.
Here, we summarize the $t$-like parameters  $(t_{(d)},\bar t_{(d)})$
and the non-factorizable terms, 
determined as
\begin{eqnarray}\label{cprime}
&&{\rm \bullet~Solution~1}\nonumber\\
&&
(t,c')=(0.25\pm0.01,0.44\pm0.01)~{\rm GeV}^3~({\rm for}~\Omega_c^0\to{\bf B}M)\,,\nonumber\\
&&
(\bar t,\bar c')=(0.52\pm0.02,0.69\pm0.01)~{\rm GeV}^3~({\rm for}~\Omega_c^0\to{\bf B}V)\,,\nonumber\\
&&
(t_d,e_{dM})=(0.16\pm0.01,0.35\pm0.01)~{\rm GeV}^3~({\rm for}~\Omega_c^0\to{\bf B}^* M)\,,\nonumber\\
&&
(\bar t_d,\bar e_{dM})=(0.26\pm0.02,0.57\pm0.06)~{\rm GeV}^3~({\rm for}~\Omega_c^0\to{\bf B}^* V)\,,
\nonumber\\
&&{\rm \bullet~Solution~2}\nonumber\\
&&
(t,c')=(0.25\pm0.01,0.06\pm0.01)~{\rm GeV}^3~({\rm for}~\Omega_c^0\to{\bf B}M)\,,\nonumber\\
&&
(\bar t,\bar c')=(0.52\pm0.02,0.35\pm0.01)~{\rm GeV}^3~({\rm for}~\Omega_c^0\to{\bf B}V)\,,\nonumber\\
&&
(t_d,e_{dM})=(0.16\pm0.01,-0.02\pm0.01)~{\rm GeV}^3~({\rm for}~\Omega_c^0\to{\bf B}^* M)\,,\nonumber\\
&&
(\bar t_d,\bar e_{dM})=(0.26\pm0.02,-0.03\pm0.02)~{\rm GeV}^3~({\rm for}~\Omega_c^0\to{\bf B}^* V)\,,
\end{eqnarray}
where the errors reflect the uncertainties of the data inputs.
Subsequently, we present two scenarios to predict the branching fractions,
denoted as $(S1,S2)$. These scenarios consider the $t$-like terms and 
two sets of the non-factorizable parameters 
based on Solution~1 and Solution~2 as outlined in Eq.~(\ref{cprime}). 
The results for these two scenarios can be found in Table~\ref{tab3}.

%
\begin{table}[b]
\caption{
Branching fractions of our work are in comparison with
those of the pole model (PM) and the data inputs.  
In ${\cal M}_{\text{STDA}}(\Omega_c^0\to{\bf B}^{(*)}M,{\bf B}^{(*)}V)$,
STDA denotes the simplified TDA, with 
$e^s_{dB(dM)}$ and $\bar e^s_{dB(dM)}$ standing for
the broken effects of the $SU(3)_f$ symmetry.}\label{tab3}
\tiny
\begin{tabular}{|l|lccc|}
\hline
Decay mode
&$\;\;\;\;\;$ ${\cal M}_{\rm{STDA}}$
&$\;\;\;$ ${\cal B}\times 10^4$ $(S1,\;S2) $$\;\;\;$
&${\cal B}\times 10^4$ (PM~\cite{Hu:2020nkg})
&$\;\;\;$ ${\cal B}\times 10^4$ (data input) $\;\;\;$\\
\hline\hline
$\Omega_c^0 \to \Xi^0 \bar{K}^{0}$
&$-(c+c^\prime)$
&$(47.0\pm7.2,47.0\pm7.2)$
&$378.0$
&$48.5\pm9.5$
\\
$\Omega_c^0 \to \Lambda^0 \bar{K}^{0}$
&$\frac{2}{\sqrt{6}}c^\prime s_c$
&$(9.0\pm0.4,0.2\pm0.1)$
&$80.5$
&\\
$\Omega_c^0 \to \Sigma^+ K^{-}$
&$0$
&$<1$
&$23.2$
&
\\
$\Omega_c^0 \to \Sigma^0 \bar{K}^{0}$
&$0$
&$<1$
&$0.9$
&
\\
$\Omega_c^0 \to \Xi^- \pi^+$
&$-ts_c$
&$(4.7\pm0.4,4.7\pm0.4)$
&$93.4$
&$4.8\pm0.5$
\\
$\Omega_c^0 \to \Xi^0 \pi^0$
&$\frac{-1}{\sqrt{2}}c s_c$
&$(2.3\pm0.2,2.3\pm0.2)$
&$54.6$
&
\\
$\Omega_c^0\to \Xi^0 \eta$
&$\frac{1}{\sqrt{6}}(3c+2c^\prime)s_c$
&$(0.2\pm0.1,4.5\pm0.5)$
&
&\\
\hline
$\Omega_c^0 \to \Xi^0 \bar{K}^{*0}$
&$-(\bar{c}+\bar{c}^\prime)$
&$(30.2\pm8.4,30.2\pm8.4)$
&
&$30.2\pm7.5$
\\
$\Omega_c^0 \to \Lambda^0 \bar{K}^{*0}$
&$\frac{2}{\sqrt{6}}\bar{c}^\prime s_c$
&$(17.7\pm0.5,4.6\pm0.3)$
&
&\\

$\Omega_c^0 \to \Sigma^+ K^{*-}$
&$0$
&$<1$
&
&\\

$\Omega_c^0 \to \Sigma^0 \bar{K}^{*0}$
&$0$
&$<1$
&
&\\
$\Omega_c^0 \to \Xi^- \rho^+$
&$-\bar{t}s_c$
&$(16.4\pm1.3,16.4\pm1.3)$
&
&\\
$\Omega_c^0 \to \Xi^0 \rho^0$
&$\frac{-1}{\sqrt{2}}\bar{c}s_c$
&$(8.2\pm0.6,8.2\pm0.6)$
&
&\\
$\Omega_c^0\to \Xi^0 \omega$
&$\frac{1}{\sqrt{2}}\bar{c}s_c$
&$(8.2\pm0.6,8.2\pm0.6)$
&
&\\
$\Omega_c^0\to \Xi^0 \phi$
&$-(\bar{c}+\bar{c}^\prime)s_c$
&$(1.4\pm0.4,1.4\pm0.4)$
&
&\\
\hline\hline
$\Omega_c^0 \to \Xi^{* 0} \bar{K}^{0}$
&$\frac{1}{\sqrt{3}}c_d$
&$(9.8\pm1.3,9.8\pm1.3)$
&
&\\
$\Omega_c^0 \to \Omega^- \pi^+$
&$t_d$
&$(28.1\pm3.6,28.1\pm3.6)$
&
&$29.6\pm2.5$
\\
$\Omega_c^0 \to \Omega ^- K^{+}$
&$(t_d-e_{dM}+\delta e^s_{dB})s_c$
&$(1.8\pm0.3,1.7\pm0.3)$
&
&$1.8\pm0.3$
\\
$\Omega_c^0 \to \Sigma^{* +}K^{-}$
&$\frac{1}{\sqrt{3}}e_{dM}s_c$
&$(2.7\pm0.2,0.01\pm0.01)$
&
&\\
$\Omega_c^0 \to \Sigma^{*0} \bar K^{0}$
&$\frac{1}{\sqrt{6}}e_{dM}s_c$
&$(1.4\pm0.1,0.004^{+0.006}_{-0.004})$
&
&\\
$\Omega_c^0 \to \Xi^{*-} \pi^+$
&$\frac{-1}{\sqrt{3}}(t_d+e_{dM})s_c$
&$(5.9\pm0.3,0.4\pm0.1)$
&
&\\
$\Omega_c^0 \to \Xi^{*0} \pi^0$
&$\frac{1}{\sqrt{6}}(c_d-e_{dM})s_c$
&$(2.9\pm0.2,0.2\pm0.1)$
&
&\\
$\Omega_c^0\to \Xi^{* 0} \eta$
&$\frac{-1}{\sqrt{2}}(c_d+e_{dM}+\frac{2}{3}\delta e^s_{dM})s_c$
&$(1.1\pm0.2,1.1\pm0.2)$
&
&\\
\hline
$\Omega_c^0 \to \Xi^{*0} \bar{K}^{*0}$
&$\frac{1}{\sqrt{3}}\bar{c}_d$
&$(18.7\pm3.0,18.7\pm3.0)$
&
&$15.1\pm5.2$
\\
$\Omega_c^0 \to \Omega^-\rho^+$
&$\bar{t}_d$
&$(46.3\pm7.4,46.3\pm7.4)$
&
&$53.3\pm10.8$
\\
$\Omega_c^0 \to \Omega^- K^{*+}$
&$(\bar{t}_d-\bar{e}_{dM}+\delta \bar{e}^s_{dB})s_c$
&$(2.5\pm1.1,1.4\pm0.4)$
&
&\\
$\Omega_c^0 \to \Sigma^{*+}K^{*-}$
&$\frac{1}{\sqrt{3}}\bar{e}_{dM}s_c$
&$(5.7\pm1.3,0.02^{+0.03}_{-0.02})$
&
&\\
$\Omega_c^0 \to \Sigma^{*0} \bar K^{*0}$
&$\frac{1}{\sqrt{6}}\bar{e}_{dM}s_c$
&$(2.8\pm0.6,0.01\pm0.01)$
&
&\\
$\Omega_c^0 \to \Xi^{*-} \rho^+$
&$\frac{-1}{\sqrt{3}}(\bar{t}_d+\bar{e}_{dM})s_c$
&$(11.7\pm1.8,1.4\pm0.3)$
&
&\\
$\Omega_c^0 \to \Xi^{*0} \rho^0$
&$\frac{1}{\sqrt{6}}(\bar{c}_d-\bar{e}_{dM})s_c$
&$(5.8\pm0.9,0.7\pm0.1)$
&
&\\
$\Omega_c^0 \to \Xi^{*0} \omega$
&$\frac{-1}{\sqrt{6}}(\bar{c}_d+\bar{e}_{dM})s_c$
&$(0.8\pm0.4,0.4\pm0.1)$
&
&\\
$\Omega_c^0 \to \Xi^{*0} \phi$
&$\frac{1}{\sqrt{3}}(\bar{c}_d+\bar{e}_{dM}+\delta \bar e^s_{dM})s_c$
&$(1.1\pm0.5,0.6\pm0.2)$
&
&\\
\hline
\end{tabular}
\end{table}
%
\section{Discussion and Conclusions}
The charm quark, not as heavy as the beauty quark, 
induces a correction to the heavy quark limit~\cite{Neubert:1993mb}. 
Consequently, the non-factorizable effect, 
negligible in beauty hadron decays~\cite{Feldmann:2004mg}, 
can be as sizeable as the factorizable effect in charm hadron decays, 
posing a challenge for estimation. 
Some studies have resorted to using $SU(3)_f$ parameters and 
global fits to extract the (non-)factorizable contribution in charmed baryon decays~\cite{Zhao:2018mov,Hsiao:2020iwc,Hsiao:2021nsc,Huang:2021aqu,Xing:2023dni}. 
Currently, the $\Omega_c^0$ decays are insufficient for a comprehensive global fit. 
In Eq.~(\ref{re3}), the approximate equivalence relations
enable us to combine or neglect certain parameters, thereby reducing their number. 
This facilitates the determination of the topological parameters, 
even in the absence of sufficient experimental data.

For $\Omega_c^0\to{\bf B}M$, 
$c^{\prime\prime}$ and $c^\prime$, as well as $e^{\prime\prime}$ and $e^\prime$, 
are identical in topology but involve different anti-symmetric quark pairs in the baryon.
According to the KPW theorem and the equivalence relation in Eq.~(\ref{re1}),
it follows that $c^{\prime\prime}=0$. Similarly, 
$e^{\prime\prime}\simeq 0$ is supported by the KPW theorem and
the approximate equivalence relations in Eq.~(\ref{re3}).
Since we demonstrate that 
$e_{\bf B}(e'_M)$ related to the 15-plet parts of the weak Hamiltonian 
contributes 35\% of its topological counterpart $e'_{\bf B}(e_M)$, 
$e_{\bf B}(e'_M)$ is neglected in the approximate equivalence relation $e_{\bf B}(e'_M)\simeq 0$.
This is similar to the common disregard for $SU(3)_f$ symmetry breaking effects, 
which are around 30\%.
We therefore conclude that although there are two parameters 
for each non-factorizable topological diagram, only one parameter remains significant.
Due to $t\simeq -c$, $t$ and $c$ are combined. Similarly,
$e'$, $e_M$ and $e'_{\bf B}$ are combined using $e'\simeq -e_M\simeq e'_{\bf B}$.
Specifically, $e'+e_M\simeq 0$ forms a unique relation applied to 
the CA and SCS decay channels, resulting in their disappearance,
while $e'_{\bf B}$ is also absent. Thus, $t$ and $c'$ are derived 
to represent the remaining contributions to the the CA and SCS amplitudes
for $\Omega_c^0\to{\bf B}M$.

For $\Omega_c^0\to{\bf B}^* M$, since $c'_d$ and $e'_d$ are
associated with the less significant $c_+H(15)$ in the effective Hamiltonian,
we consider that $c'_d$ cannot compete with $c_d$, 
and $e'_d$ cannot compete with $e_{d\bf B}$ or $e_{dM}$.
This leads to $(c'_d, e'_d)\simeq 0$ in Eq.~(\ref{re3b}).
Additionally, we combine $t_d$ and $c_d$ with $t_d\simeq -c_d$
and combine $e_{d\bf B}$ and $e_{dM}$ with $e_{d\bf B} \simeq -e_{dM}$, without neglecting them.
Consequently, we use $t_d$ and $e_{dM}$ 
to represent the four significant parameters in ${\cal M}(\Omega_c^0\to{\bf B}^* M)$.

With our analysis for $\Omega_c^0\to{\bf B}^{(*)}M$ extended to $\Omega_c^0\to{\bf B}^{(*)}V$,
we use fewer parameters to express the amplitudes, as extracted in Eq.~(\ref{cprime}).
Thus, for the first time, we explain all existing data as listed in Eq.~(\ref{cprime}).
To illustrate our findings,
we present the following ratios using ${\cal M}_{\rm STDA}$ in Table~\ref{tab3}:
\begin{eqnarray}\label{3ratios}
&&{\cal R}_1\equiv
\frac{{\cal B}(\Omega_c^0 \to \Xi^{*0} \bar{K}^{*0})}{{\cal B}(\Omega_c^0 \to \Omega^-\rho^+)}
\simeq 1/3\,, 
\nonumber\\
&&{\cal R}_2\equiv\frac{{\cal B}(\Omega_c^0 \to \Xi^- \pi^+)}{{\cal B}(\Omega_c^0 \to \Xi^0 \bar K^0)}
\simeq s_c^2/(1-c'/t)^2\,,
\nonumber\\
&&{\cal R}_3\equiv\frac{{\cal B}(\Omega_c^0 \to \Omega^- K^+)}{{\cal B}(\Omega_c^0 \to \Omega^- \pi^+)}
\simeq s_c^2(1-e_{dM}/t_d)^2\,,
\end{eqnarray}
where $s_c^2=0.05$. 
With the extracted parameters of Eq.~(\ref{cprime}) for $(S1, S2)$, 
we obtain $c'/t=(1.8,0.2)$ and $e_{dM}/t_d=(2.2,-0.1)$, 
which lead to $({\cal R}_1, {\cal R}_2, {\cal R}_3)=(0.33,0.08,0.07)$.
These results are in good agreement with 
$({\cal R}^{\rm exp}_1, {\cal R}^{\rm exp}_2, {\cal R}^{\rm exp}_3)
=(0.28\pm 0.11,0.10\pm 0.02,0.06\pm 0.01)$, 
as estimated with the experimental data in Eq.~(\ref{data1}). 
This demonstrates the applicability of the simplified topological-diagram approach 
based on the approximate equivalence relations. 
In particular, ${\cal R}_1$ is a simplified form of 
${\cal R}^{\rm TDA}_{1}=(1/3)(\bar c_d/\bar t_d+\bar c'_d/t_d)^2$ 
using $\bar c_d/\bar t_d\simeq -1$ and $\bar c'_d/t_d\simeq 0$, 
which results in ${\cal R}_1\simeq {\cal R}^{\rm exp}_1$.

Since ${\cal B}(\Omega_c^0\to {\bf B}^{(*)}M,{\bf B}^{(*)}V)$ are partially measured 
with respect to ${\cal B}(\Omega_c^0\to \Omega^-\pi^+)$, 
$\Omega_c^0\to \Omega^-\pi^+$ undoubtedly plays a key role.
Interestingly, it happens to be a pure factorizable decay channel, 
for which the factorization approach should be able to give a reliable estimation. 
However, Refs.~\cite{Xu:1992sw,Cheng:1996cs} and Ref.~\cite{Aliev:2022gxi} 
estimate ${\cal B}(\Omega_c^0 \to \Omega^- \pi^+)\sim 10^{-2}$, 
around ten and five times larger than that in Ref.~\cite{Hsiao:2020gtc}, respectively, 
showing divergent results. This is due to the $\Omega_c^0\to{\bf B}^{(*)}$ transition form factors, 
which cannot be conclusively determined from different calculations~\cite{Xu:1992sw,Cheng:1996cs, Perez-Marcial:1989sch,Cheng:1993gf,Cheng:1995fe,Pervin:2006ie,Zhao:2018zcb,Hsiao:2020gtc,Aliev:2022gxi}.

On the other hand, our approach provides an independent perspective. 
In Eq.~(\ref{cprime}), it is evident that $t=0.25$~GeV$^3$ and $\bar t=0.52$~GeV$^3$ 
are comparable to $|T|=0.23$~GeV$^3$ and $|\bar T|=0.38$~GeV$^3$ extracted 
in ${\cal B}_{3c}\to{\bf B}M$~\cite{Hsiao:2021nsc} and 
${\cal B}_{3c}\to{\bf B}V$~\cite{B3ctoBV}, respectively. 
This suggests that the factorizable contribution in $\Omega_c^0$ should be 
as significant as those in $\Lambda_c^+$ and $\Xi_c^{+(0)}$, rather than several times larger,
given that $\Omega_c^0$, $\Lambda_c^+$, and $\Xi_c^{+(0)}$ all belong to
the charmed baryon section.

To test the factorizable effect, the absolute branching fractions of 
the (nearly) pure factorizable decay channels can be useful~\cite{Hsiao:2019wyd}. 
Here, the term ``nearly'' reflects the  fact that the initial amplitudes of $\Omega_c^0 \to \Xi^- \pi^+(\Xi^- \rho^+)$ 
carry the non-factorizable parameter $e_{\bf B}(\bar e_{\bf B})$, neglected in STDA. 
Utilizing the $t$-like parameters in Eq.~(\ref{cprime}), 
we predict
\begin{eqnarray}\label{facBR}
&&
{\cal B}(\Omega_c^0 \to \Xi^- \pi^+,\Xi^- \rho^+)
=(4.7\pm0.4,16.4\pm1.3)\times 10^{-4}\,,\nonumber\\
&&
{\cal B}(\Omega_c^0 \to \Omega^- \pi^+,\Omega^- \rho^+)
=(28.1\pm3.6,46.3\pm7.4)\times 10^{-4}\,.
\end{eqnarray}

The non-factorizable effects play a key role in testing the heavy quark limit,
which can be constrained with ${\cal R}_{2,3}\simeq {\cal R}^{\rm exp}_{2,3}$ in Eq.~(\ref{3ratios}). 
As a result, the $W$-exchange parameters $c'$ and $e_{dM}$ are found to be
two times larger than the external $W$-emission terms, or 20\% and 10\% of them, 
respectively. This suggests that other theoretical studies, such as the pole model, 
might overestimate the non-factorizable contributions. For comparison, 
we predict the absolute branching fractions in scenarios 1 and 2 ($S_1,S_2$) as
\begin{eqnarray}\label{nonfacBR}
&&
{\cal B}(\Omega_c^0 \to \Lambda^0 \bar{K}^{0})
=(9.0\pm0.4,0.2\pm0.1)\times 10^{-4}\,,\nonumber\\
&&
{\cal B}(\Omega_c^0 \to \Xi^0 \bar K^0)
=(47.0\pm7.2,47.0\pm7.2)\times 10^{-4}\,.
\end{eqnarray}
In contrast, the pole model predicts 
${\cal B}_{\rm PM}(\Omega_c^0 \to \Lambda^0 \bar{K}^{0},\Xi^- \pi^+,\Xi^0 \bar K^0) 
= (80.5,93.4,378.0)$ $\times 10^{-4}$~\cite{Hu:2020nkg}, which 
leads to an inconsistency with ${\cal R}_2^{\rm BM} \simeq 2.5{\cal R}_2^{\rm exp}$. 
Additionally, it is observed that ${\cal B}(\Omega_c^0 \to \Xi^- \pi^+)$ in Eq.~(\ref{facBR}) 
is only 5\% of ${\cal B}_{\rm PM}(\Omega_c^0 \to \Xi^- \pi^+)$, 
which can be attributed to the topology, like $e_{\bf B}$, 
neglected in STDA but significantly contributing in the pole model. 
Notably, ${\cal M}(\Omega_c^0 \to \Lambda^0 \bar{K}^{0})$,
predicted to be at most $10\%$ of ${\cal B}_{\rm BM}(\Omega_c^0 \to \Lambda^0 \bar{K}^{0})$, 
can be utilized to distinguish between the two models.

The pure non-factorizable decay channel 
$\Lambda_c^+\to\Xi^0 K^+$, measured with ${\cal B}=(5.5\pm 0.7)\times 10^{-3}$~\cite{pdg},
has been used to confirm compatibility between the non-factorizable and factorizable effects
in ${\bf B}_{3c}\to {\bf B}M$. To achieve a similar experimental confirmation 
for $\Omega_c^0$ decays, we need to identify the pure non-factorizable decay channels 
and then calculate the branching fractions using the extracted parameters in Eq.~(\ref{cprime}), 
as follows:
\begin{eqnarray}\label{pureNF}
&&{\rm \bullet~Scenario~1}\nonumber\\
&&{\cal B}(\Omega_c^0 \to \Lambda^0 \bar{K}^{*0})
=(17.7\pm 0.5)\times 10^{-4}\,,\nonumber\\
&&{\cal B}(\Omega_c^0 \to \Sigma^{* +}K^{-},\Sigma^{* +}K^{*-})
=(2.7\pm0.2,5.7\pm1.3)\times 10^{-4}\,,\nonumber\\
&&{\cal B}(\Omega_c^0 \to \Sigma^{* 0}\bar K^{0},\Sigma^{* 0}\bar K^{*0})
=(1.4\pm 0.1,2.8\pm0.6)\times 10^{-4}\,,
\nonumber\\
\nonumber\\
\nonumber\\
&&{\rm \bullet~Scenario~2}\nonumber\\
&&{\cal B}(\Omega_c^0 \to \Lambda^0 \bar{K}^{*0})
=(4.6\pm 0.3)\times 10^{-4}\,,\nonumber\\
&&{\cal B}(\Omega_c^0 \to \Sigma^{* +}K^{-},\Sigma^{* +}K^{*-})
=(0.01\pm 0.01,0.02^{+0.03}_{-0.02})\times 10^{-4}\,,\nonumber\\
&&{\cal B}(\Omega_c^0 \to \Sigma^{* 0}\bar K^{0},\Sigma^{* 0}\bar K^{*0})
=(0.004^{+0.006}_{-0.004}, 0.01\pm0.01)\times 10^{-4}\,.
\end{eqnarray}
In the first scenario ($S1$),
the predicted values are as significant as those of the factorizable channels; 
in the second scenario ($S2$), only ${\cal B}(\Omega_c^0 \to \Lambda^0 \bar{K}^{*0})$ 
can be compatible with the factorizable ones. As seen in Table~\ref{tab3}, we also obtain
${\cal B}(\Omega_c^0 \to \Sigma^+ K^{(*)-})\simeq 0$ and 
${\cal B}(\Omega_c^0 \to \Sigma^0 K^{(*)0})\simeq 0$, which 
result from the simplified relation of the amplitudes in Eq.~(\ref{re3}),
$e_M+e^\prime=c^{\prime \prime}=0$ ($\bar e_M+\bar e^\prime=\bar c^{\prime \prime}=0$).
While the neglected parameters with $c_+ H(15)$ can provide a 35\% correction to the amplitudes,
in this correction, it is expected that ${\cal B}(\Omega_c^0 \to \Sigma^+ K^{(*)-})$ and 
${\cal B}(\Omega_c^0 \to \Sigma^0 K^{(*)0})$ can at most be around $10^{-5}$,
which serve to test the STDA through future measurements.

We also observe that the simplified TDA can manifest the isospin relations  in the approximate representation,
given by
\begin{eqnarray}\label{isoRE}
&&
{\cal B}(\Omega_c^0 \to \Xi^{(*)-} \pi^+)\simeq 2{\cal B}(\Omega_c^0 \to \Xi^{(*)0} \pi^0)\,,\nonumber\\
&&
{\cal B}(\Omega_c^0 \to \Xi^{(*)-} \rho^+)\simeq 2{\cal B}(\Omega_c^0 \to \Xi^{(*)0} \rho^0)\,,\nonumber\\
&&
{\cal B}(\Omega_c^0 \to \Sigma^+K^{(*)-})\simeq{\cal B}(\Omega_c^0 \to \Sigma^0\bar K^{(*)0})\simeq 0\,,\nonumber\\
&&
{\cal B}(\Omega_c^0 \to \Sigma^{* +}K^{(*)-})\simeq 2{\cal B}(\Omega_c^0 \to \Sigma^{* 0}\bar K^{(*)0})\,,
\end{eqnarray}
where the null values of ${\cal B}(\Omega_c^0 \to \Sigma^+K^-)$ and 
${\cal B}(\Omega_c^0 \to \Sigma^0\bar K^0)$ are attributed to the negligible non-factorizable effect, 
contrasting with the predictions of the pole model 
${\cal B}(\Omega_c^0 \to \Sigma^+K^-,\Sigma^0K^0)=(23.2,0.9)\times 10^{-4}$~\cite{Hu:2020nkg}.

The approximate isospin relations in Eq.~(\ref{isoRE}) correspond to $c = -t$ 
and ${\cal M}(\Omega_c^0 \to \Xi^- \pi^+, \Xi^0 \pi^0) = -(t, c/\sqrt{2})s_c$ in STDA. 
Consequently, the 10\% uncertainty of ${\cal B}(\Omega_c^0 \to \Xi^0 \pi^0) = (2.3 \pm 0.2) \times 10^{-4}$ 
can be considered as inheriting the uncertainty from ${\cal B}(\Omega_c^0 \to \Xi^- \pi^+)$, 
where $(4.8 \pm 0.5) \times 10^{-4}$ is used as a data input. 
Additionally, the $c_+$ correction at ${\cal O}(\delta_c^1)$ may introduce a further 35\% uncertainty 
for ${\cal M}(\Omega_c^0 \to \Xi^0 \pi^0)$. 
Thus, if ${\cal B}(\Omega_c^0 \to \Xi^0 \pi^0)$ is measured 
to deviate from ${\cal B}(\Omega_c^0 \to \Xi^- \pi^+)/2$ by a certain margin, 
the applicability of STDA may be called into question.

To test the applicability of STDA,
a perspective on the $c_+$ corrections can be considered.
Due to the abundance of experimental results, approximations 
related to $\delta_c \simeq 35\%$ 
were not necessary in the TDA study for ${\bf B}_{3c} \to {\bf B}M$~\cite{Hsiao:2021nsc}. 
In this case, determining $(C,T) = (-0.23 \pm 0.02, 0.24 \pm 0.02)$~GeV$^3$ 
results in $C/T = -(0.96 \pm 0.08)$, which is consistent with 
the approximation $c = -t$ in our study. 
This correspondence suggests that STDA provides 
a reasonable approximation in charmed baryon decays 
and implies that $c_+$ corrections may not significantly alter the branching fractions.

While STDA is applied, it remains challenging to determine 
whether discrepancies with the data stem from broken $SU(3)_f$ symmetry or neglected 15-plet parameters. 
Notably, $g\to s\bar s$ in the $W$-annihilation and $W$-exchange diagrams 
plays a crucial role in breaking $SU(3)_f$ symmetry in $D\to MM$ decays \cite{Li:2012cfa,Cheng:2019ggx}. 
Similar effects are also suggested in ${\bf B}_{3c}\to {\bf B}M$ \cite{Hsiao:2020iwc,Hsiao:2021nsc}. 
For instance, the ratio ${\cal B}(\Xi_c^0\to \Xi^- K^+)/{\cal B}(\Xi_c^0\to \Xi^- \pi^+)$,
expressed as $s_c^2(2T-E_{{\bf B}}^{(s)})^2/(2T-E_{{\bf B}})^2$, 
requires $|E_{{\bf B}}^{(s)}| \simeq 1.7 |E_{{\bf B}}|$ to reconcile with the data, 
indicating significant $SU(3)_f$ breaking effects. 
Therefore, we anticipate that $e^{(s)}$ can help in observing $SU(3)_f$ breaking in $\Omega_c^0$ decays.

In Table~\ref{tab1}, several potential $e^{(s)}$ terms are identified that could induce flavor symmetry breaking 
in $\Omega_c^0\to {\bf B}^{(*)}M,{\bf B}^{(*)}V$. Following reduction in STDA, 
only $e^s_{dB(dM)}$ and $\bar e^s_{dB(dM)}$ remain, expressed as:
\begin{eqnarray}
e^s_{dB(dM)}=e_{dB(dM)}+\delta e^s_{dB(dM)}\,,\nonumber\\
\bar e^s_{dB(dM)}=\bar e_{dB(dM)}+\delta \bar e^s_{dB(dM)}\,.
\end{eqnarray}
We assume $\delta e^s_{dB,dM}=\delta \bar e^s_{dB,dM}=0$ for exact $SU(3)_f$ calculations, 
as current experimental information does not confirm 
non-zero $\delta e^s_{dB(dM)}$ or $\delta \bar e^s_{dB(dM)}$. 
We propose $\Omega_c^0 \to \Omega ^- K^{(*)+}$, $\Omega_c^0\to \Xi^{ 0} \eta$, 
and $\Omega_c^0 \to \Xi^{*0} \phi$ for future scrutiny, 
as their amplitudes could potentially reveal the breaking terms $\delta e^s_{dB(dM)}$ and 
$\delta \bar e^s_{dB(dM)}$, as outlined in Table~\ref{tab3}. Nonetheless,
the challenge with the application of STDA remains. For possible clarification, we derive the relation:
\begin{eqnarray}\label{BKrelation}
\frac{{\cal B}(\Omega_c^0 \to \Xi^{*0} \phi)}{{\cal B}(\Omega_c^0 \to \Xi^{*0} \omega)}
=\frac{2(\bar{c}_d+\bar{e}_{dM}+\bar{c}_d\delta_{V_2}+\delta \bar e^s_{dM})^2 F_{\phi}}
{(\bar{c}_d+\bar{e}_{dM})^2 F_{\omega}}\,, 
\end{eqnarray}
where $F_{\phi,\omega}$ as the dynamical factors in Eq.~(\ref{p_space}) 
lead to $F_\phi/F_\omega\simeq 0.7$. Here, 
$\bar{c}_d\delta_{V_2}$ presents a tiny breaking effect from vector meson production,
and the neglected 15-plet parameters are unlikely to influence this relation.
Consequently, $\delta \bar e^s_{dM}$ could account for the observed breaking effect,
if future measurements indicate a significant deviation from the expected value of approximately 1.4 
when $\delta \bar e^s_{dM} = 0$ in Eq.~(\ref{BKrelation}).

Our approach relies on $SU(3)_f$ symmetry, which may lose validity due to
significant breaking effects. Additionally, broken symmetry can enhance 
$CP$ asymmetry~\cite{Li:2012cfa,Cheng:2019ggx,Feldmann:2012js,Brod:2012ud,Muller:2015rna},
as observed in $D\to MM$ decays~\cite{LHCb:2016csn}. Therefore, exploring flavor symmetry breaking 
is crucial for testing the validity of our approach and useful for studying $CP$ violation 
in $\Omega_c^0$ decays.

As the final remark, 
the Cabibbo-allowed decay channels are expected to be more accessible for detection, 
such as $\Omega_c^0 \to \Xi^0 \bar{K}^{(*)0},\Xi^{*0} \bar{K}^{(*)0}$ and 
$\Omega_c^0 \to \Omega^- \pi^+(\rho^+)$, with estimated branching fractions 
at the level of $10^{-3}$, as listed in Table~\ref{tab3}. Particularly, 
$\Omega_c^0 \to \Xi^{ 0} \bar{K}^{0}$, 
as the only Cabibbo-allowed decay not yet measured, 
has a branching fraction of ${\cal B}(\Omega_c^0 \to \Xi^{* 0} \bar{K}^{0})=(9.8\pm1.3)\times 10^{-4}$, 
which could be significant for measurement. Since $e'_{\bf B}$ is specifically introduced 
for the DCS decay channels, such as $\Omega_c^0 \to \Sigma^-\pi^+$, 
it is neither negligible with the STDA relations in Eq.~(\ref{re3}) 
nor constrainable, as none of the DCS decays have been observed yet. Therefore,
we must await future measurements, such as ${\cal B}(\Omega_c^0 \to \Sigma^-\pi^+)$,
to extract $e'_{\bf B}$. This will enable a more systematic analysis of the DCS decay channels.

In summary, we have developed the simplified topological-diagram approach 
to investigate two-body non-leptonic $\Omega_c^0$ decays. 
Employing the $SU(3)_f$-induced topological approach, 
we have depicted and parameterized the $W$-emission and 
$W$-exchange processes, enabling us to establish 
stringent $SU(3)_f$ relations for potential decay channels.
To address the issue of various non-factorizable terms, 
we have utilized the irreducible $SU(3)_f$ approach,
identifying the dominant terms as $c'(\bar c')$ for 
$\Omega_c^0\to{\bf B}M(V)$ and $e_{dM}(\bar e_{dM})$ for $\Omega_c^0\to{\bf B}^*M(V)$.
With the topological parameters determined using available data, we have interpreted
the following experimental ratios:
${\cal B}(\Omega_c^0\to\Xi^{*0}\bar K^{*0})/{\cal B}(\Omega_c^0\to\Omega^-\rho^+)=0.28\pm 0.11$,
${\cal B}(\Omega_c^0\to\Xi^-\pi^+)/{\cal B}(\Omega_c^0\to\Xi^{0}\bar K^{0})=0.10\pm 0.02$, and
${\cal B}(\Omega_c^0 \to \Omega^- K^+)/{\cal B}(\Omega_c^0 \to \Omega^- \pi^+)=0.06\pm 0.01$.
Additionally, we have calculated the branching fractions for the Cabibbo-allowed decays,
such as ${\cal B}(\Omega_c^0 \to \Xi^{* 0} \bar{K}^{0})=(9.8\pm1.3)\times 10^{-4}$.
Of particular interest, we have derived approximate isospin relations:
${\cal B}(\Omega_c^0 \to \Xi^{(*)-} \pi^+)\simeq 2{\cal B}(\Omega_c^0 \to \Xi^{(*)0} \pi^0)$ and
${\cal B}(\Omega_c^0 \to \Xi^{(*)-} \rho^+)\simeq 2{\cal B}(\Omega_c^0 \to \Xi^{(*)0} \rho^0)$.
In addition, 
we have presented $\Omega_c^0 \to \Omega ^- K^{(*)+}$, $\Omega_c^0\to \Xi^{ 0} \eta$, 
and $\Omega_c^0 \to \Xi^{*0} \phi$ as potential candidates for testing $SU(3)_f$ symmetry breaking.
As a highlight,
we have predicted ${\cal B}(\Omega_c^0 \to \Xi^0 \pi^0)=(2.3\pm0.2)\times 10^{-4}$,
accessible to experimental facilities such as Belle and LHCb.

\newpage
\section*{Acknowledgments}
The authors would like to thank 
Prof.~Jinlin~Fu, Prof.~Xiao-Rui~Lyu, and Prof.~Chengping~Shen for useful discussions.
This work was supported in part
by National Science Foundation of China (Grants No.~12175128 and No.~11675030)
and Innovation Project of Graduate Education in Shanxi Province (2023KY430).


\end{document}